\documentclass[transmag]{IEEEtran}
\usepackage{latexsym}
\usepackage{graphicx}
\usepackage{amsfonts,amssymb,amsmath}
\usepackage{hyperref}
\usepackage{indentfirst,flushend}
\usepackage{amssymb,bm,mathrsfs,times,amscd,amsmath,amsthm,amsfonts,bbm}
\usepackage{latexsym}
\usepackage{dsfont,slashbox}
\usepackage{graphicx}
\usepackage{subfigure}
\usepackage{color}
\usepackage{cases}
\usepackage[justification=centering]{caption}

\newtheorem{remark}{Remark}
\usepackage{multirow}
\usepackage[sort]{cite}
\usepackage{multicol}
\usepackage{clrscode}
\usepackage{algorithm}
\usepackage{psfrag,stfloats,nicefrac}
\usepackage{amsmath}
\usepackage{amssymb}

\usepackage{algorithmic}
\usepackage{supertabular}
\usepackage{makecell}
\usepackage{epstopdf}
\def\BibTeX{{\rm B\kern-.05em{\sc i\kern-.025em b}\kern-.08em T\kern-.1667em\lower.7ex\hbox{E}\kern-.125emX}}
\begin{document}

\title{An Overview on the Application of Graph Neural Networks in Wireless Networks}

\author{Shiwen~He,~\IEEEmembership{Member,~IEEE},~Shaowen Xiong,~Yeyu Ou,~Jian Zhang~\IEEEmembership{Member,~IEEE}, Jiaheng~Wang,\\~\IEEEmembership{Senior Member,~IEEE}, Yongming~Huang,~\IEEEmembership{Senior Member,~IEEE}, and Yaoxue~Zhang,~\IEEEmembership{Senior Member,~IEEE}
\thanks{Manuscript received Jul. 29, 2021; revised Sept. 15, 2021; Oct. 25, 2021; accepted Nov. 12, 2021. The associate editor coordinating the review of this paper and approving it for publication was Prof. Pablo Serrano. This work was supported in part by National Natural Science Foundation of China under Grants 62171474 and 61720106003, in part by the open research fund of National Mobile Communications Research Laboratory Southeast University under Grants No. 2022D03, in part by the Natural Science Foundation of Hunan Province under Grants 2020JJ4745.(\emph{Corresponding author: Jian Zhang;Yongming Huang})}
\thanks{S. He, S. Xiong, Y. Ou, and J. Zhang are with the School of Computer, Central South University, Changsha 410083, China. S. He is also with the National Mobile Communications Research Laboratory, Southeast University, and the Purple Mountain Laboratories, Nanjing 210096, China. (email: \{shiwen.he.hn, shaowen.xiong, ouyeyu, jianzhang\}@csu.edu.cn).}
\thanks{J. Wang and Y. Huang are with the National Mobile Communications Research Laboratory, School of Information Science and Engineering, Southeast University, Nanjing 210096, China. They are also with the Purple Mountain Laboratories, Nanjing 210096, China. (email: \{jhwang, huangym\}@seu.edu.cn). }
\thanks{Y. Zhang is with the Department of Computer Science and Technology, Tsinghua University, Beijing 100084, China. (email: zhangyx@tsinghua.edu.cn)}
}

\IEEEtitleabstractindextext{\begin{abstract}In recent years, with the rapid enhancement of computing power, deep learning methods have been widely applied in wireless networks and achieved impressive performance. To effectively exploit the information of graph-structured data as well as contextual information, graph neural networks~(GNNs) have been introduced to address a series of optimization problems of wireless networks. In this overview, we first illustrate the construction method of wireless communication graph for various wireless networks and simply introduce the progress of several classical paradigms of GNNs. Then, several applications of GNNs in wireless networks such as resource allocation and several emerging fields, are discussed in detail. Finally, some research trends about the applications of GNNs in wireless communication systems are discussed.\end{abstract}


}

\maketitle

\section{INTRODUCTION}
\IEEEPARstart{T}{he} advent of fifth-generation~(5G) wireless communication systems has driven the revolutionary applications extending far beyond smartphones and other mobile devices~\cite{HePreceeding2021}. Meanwhile, intelligent communication becomes a novel developing trend of future communication systems~\cite{DLwireless2018}. Recently, more and more researchers adopt the deep learning~(DL) method to study the problems in wireless networks motivating by the successful application of DL in the related fields of computer~\cite{DLmobilewireless2018}.

\subsection{From Traditional Deep Learning to Graph Neural Network}
According to the usage of domain knowledge, DL methods can be divided into the data-driven DL method and the data- and model-driven DL method. The data-driven DL methods without using the domain knowledge generally have poor interpretability and robustness, while having fast inferencing speed compared with the model-based method. A classic work using multi-layer perceptrons~(MLPs) is to solve the power control problem of wireless networks by using the MLPs to approximate the weighted minimum mean square error~(WMMSE) algorithm~\cite{learnOpt2018}. The data- and model-driven DL methods are rising in recent years and its core idea is to retain the main theoretical characteristics of the classical model algorithms, while using the DL methods to partially enhance or replace its related difficult or time-consuming process. Compared with the data-driven DL methods, the data- and model-driven DL methods have better interpretability and robustness with slower inferencing speed. For example, A. Bora~\emph{et al.} used the generated model from neural networks instead of the standard sparsity model to represent data distributions~\cite{CSGM2017}. N. Shlezinger~\emph{et al.} designed a deep neural network~(DNN) to implement the channel-dependent part of Viterbi algorithm, while keeping the unchange of the rest remains~\cite{ViterbiNet2020}. H. Kim~\emph{et. al} studied a family of sequential codes parameterized by recurrent neural network~(RNN) architectures~\cite{kim2018communication}. K. Gregor~\emph{et al.} designed a non-linear and parameterized feed-forward architecture with a fixed depth to approximate the optimal sparse code~\cite{LISTA2010}. Some overviews are presented to summary the application of machine learning~(ML) or DL with aiming to improve the quality-of-experience~(QoE) of wireless networks~\cite{MLwirelessLayerSurvey2021, ML6GSurvey2021}. The data collected in these tasks is typically represented in the Euclidean domains. Although the existing works using the DL models defined in the Euclidean domains achieve a better performance in small-scale networks, they fail to exploit the underlying topology of wireless networks. Consequently, the performance decreases sharply when the network scale becomes large~\cite{OptPowerContrl2020,PowerContrlCNN2018}.

In wireless networks, an obvious feature is that the high dynamics of network topology caused by some uncertain factors, e.g., the user mobility, changes in traffic pattern or adjustment of the network resource, etc. In addition, the wireless data may be collected from non-Euclidean domains and represented as graph-structured data with high dimensional features and interdependency among communication devices. These issues bring difficulties to apply directly the learning model defined in Euclidean domains in wireless networks. A straightforward way solving these difficulties is to incorporate the network topology information, which is described as an adjacency matrix depending on the specific node index, into the architecture of neural networks. However, the indices of communication devices in wireless networks may change due to the reallocation of resources and the movement of communication devices, i.e., dynamic graph-structured data. This motivates us to design a novel learning model with taking into account the interdependencies between communication devices and the dynamics of wireless networks~\cite{zhang2021topology}. The emerging GNNs enable the graph-structured data to be processed effectively and to use the global parameterization, common system of coordinates, vector space structure, or shift-invariance~\cite{GeoDL2017}. In the last few years, many researchers have begun to use GNNs to mine the deep information hidden in the graph-structured data to further improve the abilities of learning and simulating the interaction between nodes.
\begin{figure*}
\centerline{\includegraphics[width=1.8\columnwidth,keepaspectratio]{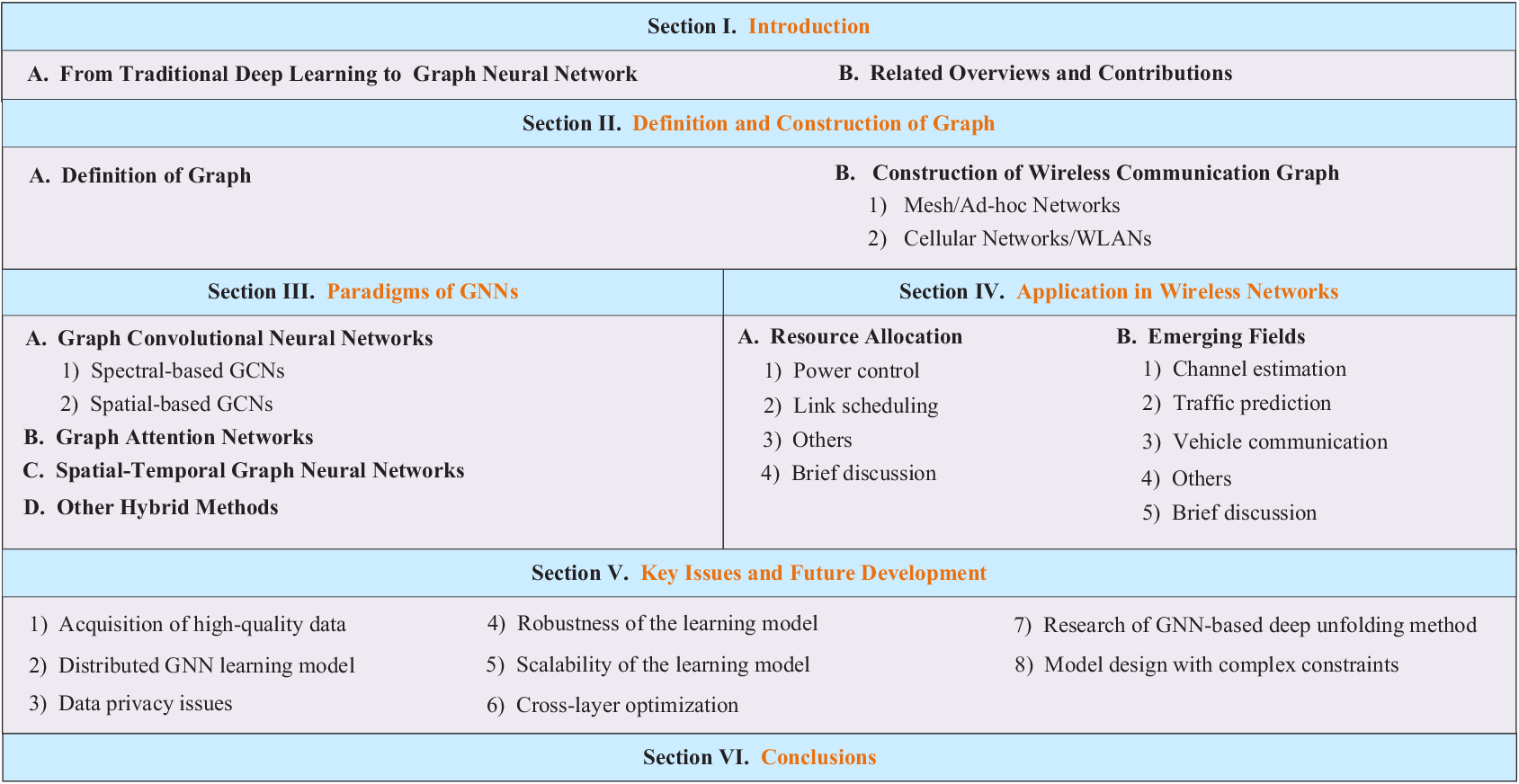}}
\caption{Road map of the overview.\label{roadmap}}
\end{figure*}

\subsection{Related Overviews and Contributions}
Some overviews about the paradigms and applications of GNNs are represented in the past few years. The authors of~\cite{GNNsurvey12020} introduced comprehensively four basic paradigms of GNNs and described the representative models in detail. Differentiating from~\cite{GNNsurvey12020}, the authors of~\cite{graphDLSurvey2020} further reviewed graph reinforcement learning and graph adversarial methods. The authors of~\cite{xia2021graph} summarized the state-of-the-art of the main models and algorithms of graph learning from four aspects, i.e., graph signal processing, matrix decomposition, random walk, and deep learning. The aforementioned surveys introduced in detail the characteristics of different GNN paradigms but briefly introduced the application of GNNs in some fields except for wireless networks. The authors of~\cite{SurveySpatialSpectral2020} summarized the GNN approaches in terms of the spatial domain and spectral domain, respectively. Furthermore, this work integrates the spatial and spectral domain models into a unified framework. The authors of~\cite{SurveyGNNexpressPower2020} further discussed the expressive power of GNNs and further summarized the relationships among GNNs, the Weisfeiler-Lehman algorithm, and distributed local algorithms. The authors of~\cite{ruiz2021graph} illustrated the excellent performance of GNNs depending on the three characteristics of equivariance, stability and transferability, which are further determined by the selection of the optimization objective and technologies, as well as the design of graph filters. The methods using software or hardware to accelerate GNNs were reviewed from the perspective of computation speed in~\cite{computingGNNsurvey2020}. In order to fill the research on the interpretability of GNNs, the unified methodology and standard testbed for evaluating the interpretability of GNNs were summarized comprehensively in~\cite{ExplainGNNsurvey2020}. There are also several overviews on applying the GNNs to solve the problems in the traffic domain~\cite{graphTrafficDomain2020}, power systems~\cite{graphPowerSystem2021}, and recommender system~\cite{graphKGRS2021}. In addition, the relationship between GNNs and the latest neural-symbolic computing that aims at integrating the abilities of learning from the environment and of reasoning from what has been learned was introduced in~\cite{GNNNeuSymbol2020}.

In this paper, we aim to present a comprehensive overview of the application of GNNs in wireless networks. Meanwhile, we also provide some potential research directions for researchers who are interested in this topic. A detailed organization of this overview is illustrated in Fig.~\ref{roadmap}. In particular, the main contributions of this overview are summarized as follow
\begin{itemize}
  \item The methods of constructing wireless communication graph~(WCG) for Mesh/Ad-hoc networks, Cellular networks, or Wireless Local Area Networks~(WLANs) are illustrated elaborately.
  \item Several classical paradigms of GNNs applied in wireless networks are introduced to acquire a better understanding of the concepts and structures of GNNs.
  \item A comprehensive review of GNNs applied in wireless networks is summarized in terms of the existing directions, e.g., resource allocation and several emerging fields.
  \item Some challenges and potential research directions are summarized and discussed for the application of GNNs in wireless networks.
\end{itemize}

The rest of this paper is organized as follows: In Section~\uppercase\expandafter{\romannumeral2}, we introduce several basic definitions of graph-structured data and summarize the construction methods of WCG for Mesh/Ad-hoc networks and Cellular networks/WALNs. In Section~\uppercase\expandafter{\romannumeral3}, we review several classical paradigms of GNNs that applied in wireless networks. In Section~\uppercase\expandafter{\romannumeral4}, we introduce the application of GNNs in wireless networks. In Section~\uppercase\expandafter{\romannumeral5}, we discuss a few valuable directions for the application of GNNs in wireless networks. Finally, we conclude this work in Section~\uppercase\expandafter{\romannumeral6}. For ease of reading, the notations commonly used in graphs are summarized in Table~\uppercase\expandafter{\romannumeral1}.
\begin{table}[h!t]
\centering
\caption{\textsc{Commonly Used Notations}}
\begin{tabular}{|c|l|}
	\cline{1-2}
	$G$       & Graph                                                                 \\ \cline{1-2}
    $N$    & The number of nodes                                               \\ \cline{1-2}
	$V$       & The set of nodes                                                             \\ \cline{1-2}
	$v$   & A node $v\in V$                                \\ \cline{1-2}
	$\mathcal{N}(v)$    & The set of neighbor nodes of $v\in V$                               \\ \cline{1-2}
	$E$       & The set of edges                                                           \\ \cline{1-2}
	$e_{ij}$  & An edge from node $v_i$ to node $v_j$                 \\ \cline{1-2}
	$\mathbf{A}$    & Adjacent matrix                                \\ \cline{1-2}
    $\mathbf{A}^T$    & The transpose of matrix $\mathbf{A}$                                \\ \cline{1-2}
    $\mathbf{A}_{ij}$    & \makecell[l]{The element of the $i$-th row and the $j$-th \\column of matrix $\mathbf{A}$}  \\ \cline{1-2}
	$\mathbf{D}$    & The degree matrix of matrix $A$ \\ \cline{1-2}
    $\mathbf{D}^{-\frac{1}{2}}$    & $-\frac{1}{2}$ power of the elements in the matrix $\mathbf{D}$  \\ \cline{1-2}
	$\mathbf{X}$    & Feature matrix                                    \\ \cline{1-2}
	$\mathbf{X}_{i}$   & The $i$-th column of matrix $\mathbf{X}$                      \\ \cline{1-2}
    $\mathbf{x}[l]$   &  Denotes the $l$-th element of vector $\mathbf{x}$ \\ \cline{1-2}
	$d$    & The dimension of the feature vector                                          \\ \cline{1-2}
    $L$    & The maximum depth of GNNs \\ \hline
\end{tabular}
\end{table}

\section{Definition and Construction of Graph}
In this section, we would like to first describe several basic definitions of graph-structured data. Then, the construction methods of WCG for different wireless network scenarios, e.g., Mesh/Ad-hoc networks, Cellular networks, and WLANs, are illustrated in detail.

\subsection{Definition of Graph}
Graph-structured data is a kind of non-Euclidean data and is commonly expressed as $G = (V, E)$, where $V$ and $E$ are the sets of nodes and edges, respectively~\cite{GNNsurvey2019}. Let $v_i\in V$ be a node and $e_{ij}=(v_i, v_j)\in E$ be an edge from node $v_i$ to node $v_j$. The adjacency matrix of a graph is represented as $\mathbf{A}$. If $e_{ij}\in E$, $\mathbf{A}_{ij}=1$, otherwise, $\mathbf{A}_{ij} =0$. The graph is undirected if $\mathbf{A}$ is symmetric, otherwise, the graph is directed. The degree matrix $\mathbf{D}$ of an undirected graph is a diagonal matrix, where $\mathbf{D}_{ii}=|\mathcal{N}(v_i)|$. The Laplacian matrix of an undirected graph is defined as $\mathbf{L}=\mathbf{D}-\mathbf{A}$. The normalized Laplacian matrix is defined as $\tilde{\mathbf{L}}=\mathbf{I}_N-\mathbf{D}^{-\frac{1}{2}}\mathbf{AD}^{-\frac{1}{2}}$. Note that the normalized Laplacian matrix $\tilde{\mathbf{L}}$ is a real semi-positive definite matrix. Accordingly, it can be decomposed into $\tilde{\mathbf{L}}=\mathbf{U}\mathbf{\Lambda}\mathbf{U}^{T}$, where $\mathbf{U}$ is the eigenvector matrix and $\mathbf{\Lambda}$ is a diagonal matrix of $[\mathbf{\Lambda}]_{ii}=\lambda_{i}$ with $\lambda_{i}$ being the eigenvalue. While for the directed graph, the in-degree and the out-degree matrices are defined as $\mathbf{D}_{jj}^{(in)}=\sum_{i}^{N}\mathbf{A}_{ij}$ and $\mathbf{D}_{ii}^{(out)}=\sum_{j}^{N}\mathbf{A}_{ij}$, respectively. The transition probability matrix $\mathbf{P}$ of a given directed graph is defined as $\mathbf{P}=\left(\mathbf{D}^{(out)}\right)^{-1}\mathbf{A}$. Accordingly, a symmetric normalized Laplacian of directed graph is defined as $\overrightarrow{\mathbf{L}}=\mathbf{I}-\frac{1}{2}\left(\mathbf{\Phi}^{\frac{1}{2}}\mathbf{P}\mathbf{\Phi}^{-\frac{1}{2}}+\mathbf{\Phi}^{-\frac{1}{2}}\mathbf{P}^T\mathbf{\Phi}^{\frac{1}{2}}\right)$, where $\mathbf{\Phi}$ is generated according to $\mathbf{P}$ and perron vector~\cite{ma2019spectral}. It is worth mentioning that, from the reviewing results, there is almost no work to deal with directed graph based on spectral domain, but based on spatial domain. In a graph, each node may have its own attribute feature. The feature matrix of a graph is defined as $\mathbf{X}\in\mathbb{R}^{N\times{d}}$. If feature matrix $\mathbf{X}$ changes over time, the graph is defined as a spatial-temporal graph.

\subsection*{B. Construction of Wireless Communication Graph}
The first thing of using GNNs is to transform a wireless network into a graph. In general, according to the specific goal of the research, the topology structure of wireless networks may be constructed into an undirected graph or directed graph. According to the types of communication links and communication devices in wireless networks, it can be further constructed into a homogeneous or heterogeneous graph. In the sequel, the construction methods of WCG for various wireless network scenarios, e.g., Mesh/Ad-hoc networks, Cellular networks, or WLANs, are illustrated elaborately.

\setcounter{subsubsection}{0}
\subsubsection{Mesh/Ad-hoc Networks} 
\begin{figure}[htbp]
\renewcommand{\captionfont}{\footnotesize}
	\centering
\includegraphics[width=1.0\columnwidth,keepaspectratio]{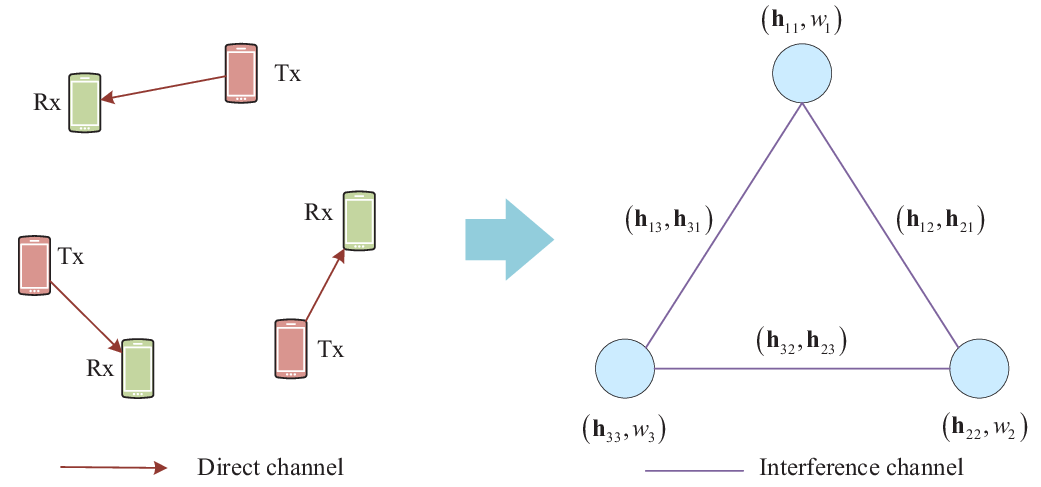}\\
	\caption{Illustration of constructing WCG for homogeneous Mesh/Ad-hoc networks with 3 transceivers~\cite{gnnWPC2019}.}
	\label{K-user-interference}
\end{figure}

Suppose there are $N$ transceiver pairs in homogeneous Mesh/Ad-hoc networks. To build a WCG for this kind of wireless network, we view the $i$-th transceiver pair as the $i$-th node of WCG, the feature vector of the $i$-th node includes the direct channel state information~(CSI) $\mathbf{h}_{ii}$~\footnote{It is worth noting that the dimension of $\mathbf{h}_{ij}$ between nodes $v_i$ and $v_j$ is determined by the number of antennas equipped by the transceiver pair.} and other environmental information, such as the weight $w_i$ of the $i$-th node. The edge between nodes $v_i$ and $v_j$ in WCG may be undirected or directed. The feature vector of the undirected edge includes the interference CSIs $\mathbf{h}_{ij}$ and $\mathbf{h}_{ji}$. While the feature vectors of two directed edges between nodes $v_i$ and $v_j$ can include $\mathbf{h}_{ij}$ and $\mathbf{h}_{ji}$, respectively~\footnote{The construction method of the feature vectors of nodes and edges is not limited in the method aforementioned and mentioned later, which can be adjusted and supplied according to the specific research tasks.}. Fig.~\ref{K-user-interference} shows a construction method of WCG for homogeneous Mesh/Ad-hoc networks with 3 transceivers.

\begin{figure}[htbp]
\renewcommand{\captionfont}{\footnotesize}
	\centering
\includegraphics[width=1.0\columnwidth,keepaspectratio]{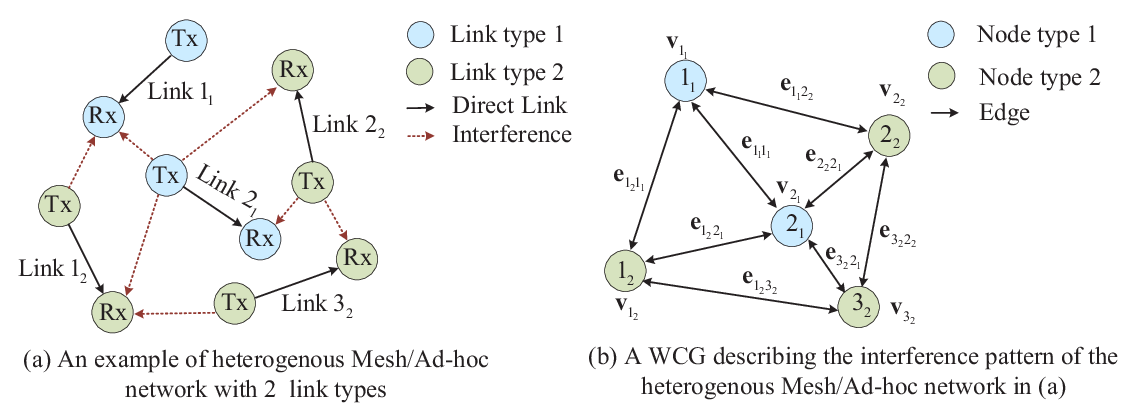} \\
	\caption{Illustration of constructing WCG for heterogeneous Mesh/Ad-hoc networks with 2 link types~\cite{PowerContrlBF2021}.}
	\label{heterograph}
\end{figure}

In heterogeneous Mesh/Ad-hoc networks, suppose there are $N$ types of communication links. We treat the $i$-th transceiver pair with link type $m$ as node $v_{i_m}$ in the WCG. The set of neighboring nodes with link type $n$ of node $v_{i_m}$ are represented as $\mathcal{N}_{i_m}^{(n)}$. The feature vector of node $v_{i_m}$ includes the CSI $\mathbf{h}_{i_mi_m}$ of direct link, and other environmental information corresponding to the link of type $m$. The feature vector of the edge between nodes $v_{i_m}$ and $v_{j_n}$ should also be considered from the perspective of undirected and directed. The feature vector of the undirected edge includes the CSIs, i.e., $\mathbf{h}_{i_mj_n}$ and $\mathbf{h}_{j_ni_m}$ of interference links. While the feature vectors of the directed edges between nodes $v_{i_m}$ and $v_{j_n}$ include the $\mathbf{h}_{i_mj_n}$ and $\mathbf{h}_{j_ni_m}$, respectively. Fig.~\ref{heterograph} illustrates a construction method of WCG for heterogeneous Mesh/Ad-hoc networks with 2 link types, where $\mathbf{v}_{j_n}$ is the feature vector of node $v_{j_n}$, $\mathbf{e}_{j_ni_m}$ is the feature vector of the edge between nodes $v_{j_n}$ and $v_{i_m}$, and ``Link $i_m$" indicates the $i$-th communication link with type $m$.

\subsubsection{Cellular Networks/WLANs} 
\begin{figure}[htbp]
\renewcommand{\captionfont}{\footnotesize}
	\centering
\includegraphics[width=1.0\columnwidth,keepaspectratio]{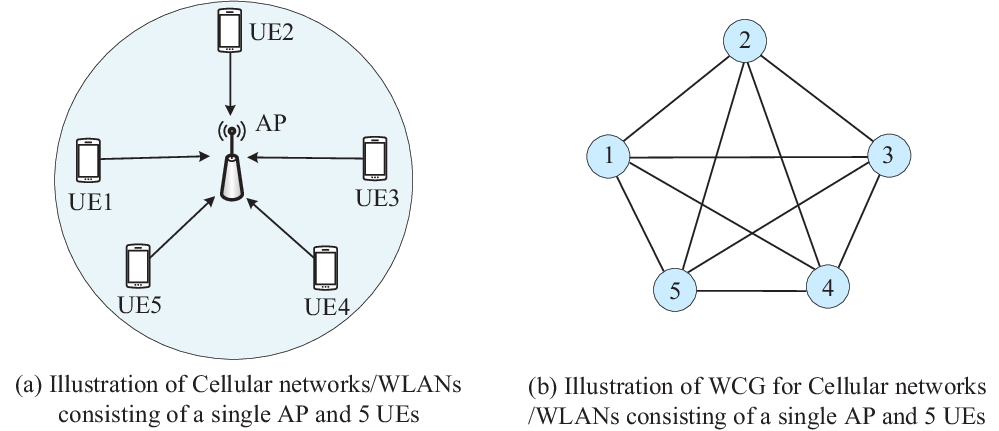} \\
	\caption{Illustration of constructing WCG for Cellular networks/WLANs with a single AP.}
	\label{CellularSingleAP}
\end{figure}

In general, Cellular networks/WLANs may consist of $M$~($M\geq 1$) access points~(APs) and $N$~($N\geq 1$) user equipments~(UEs). Considering a simple situation including only one AP, which allocates the resources, such as power control and user association, etc., to the UEs. We treat the $i$-th UE as a node of WCG should be built while ignoring AP. The feature vector of the $i$-th node includes the CSI $\mathbf{h}_{ii}$ and other environmental information. The feature vector of the edge between nodes $v_i$ and $v_j$ includes the CSI $\mathbf{h}_{ii}$ and $\mathbf{h}_{jj}$, etc., which can also be ignored due to all UEs share one AP. An illustration of constructing WCG for Cellular networks/WLANs with a single AP is shown in Fig.~\ref{CellularSingleAP}.

\begin{figure}[htbp]
\renewcommand{\captionfont}{\footnotesize}
	\centering
\includegraphics[width=1.0\columnwidth,keepaspectratio]{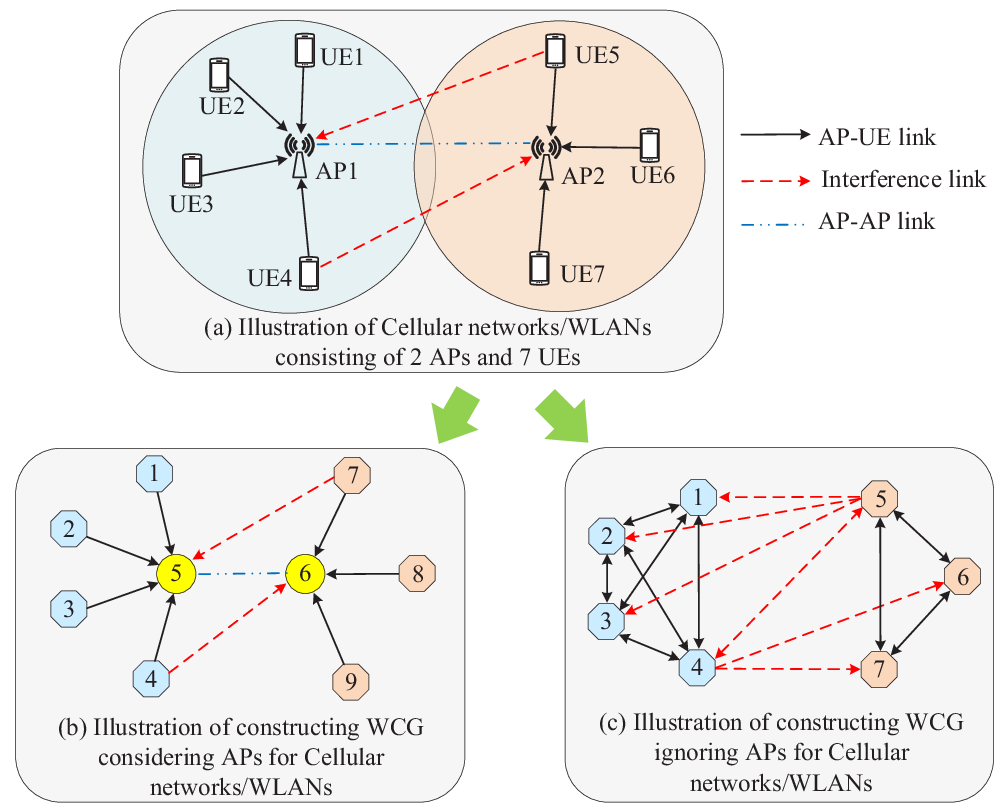} \\
	\caption{Illustration of constructing WCG for Cellular networks/WLANs with multiple APs.}
	\label{CellularMultiAPs}
\end{figure}

We further consider a more complex situation including multiple APs in Cellular Networks/WLANs, in which one AP may serve multiple UEs, and one UE may also access multiple APs. We first consider the scenario that one UE just accesses one AP and one AP serves multiple UEs. As shown in Fig.~\ref{CellularMultiAPs}~(a), UEs 1-4 and UEs 5-7 communicate with AP1 and AP2, while AP2 and AP1 are interfered with UE4 and UE5, respectively. The WCG of this kind of wireless networks can be built in two manners, which are illustrated in Fig.~\ref{CellularMultiAPs}~(b) and Fig.~\ref{CellularMultiAPs}~(c), respectively. In Fig.~\ref{CellularMultiAPs}~(b), there are two types of nodes indicating APs and UEs, respectively. The information of position, channel configuration, and device type are considered to be the feature vector of nodes. The feature vector of the edge between nodes $v_i$ and $v_j$ includes the direct/interference CSI and other link information. In contrast to Fig.~\ref{CellularMultiAPs}~(b), the APs are ignored in Fig.~\ref{CellularMultiAPs}~(c), which includes only one type of nodes indicating UE. The feature vector of a node includes the position of UE, channel configuration and device type, etc. The direct/interference CSIs and other link information can be considered to be the feature vector of edge. In practice, we can choose the appropriate manner according to the specific research tasks.

\begin{figure}[htbp]
\renewcommand{\captionfont}{\footnotesize}
	\centering
\includegraphics[width=1.0\columnwidth,keepaspectratio]{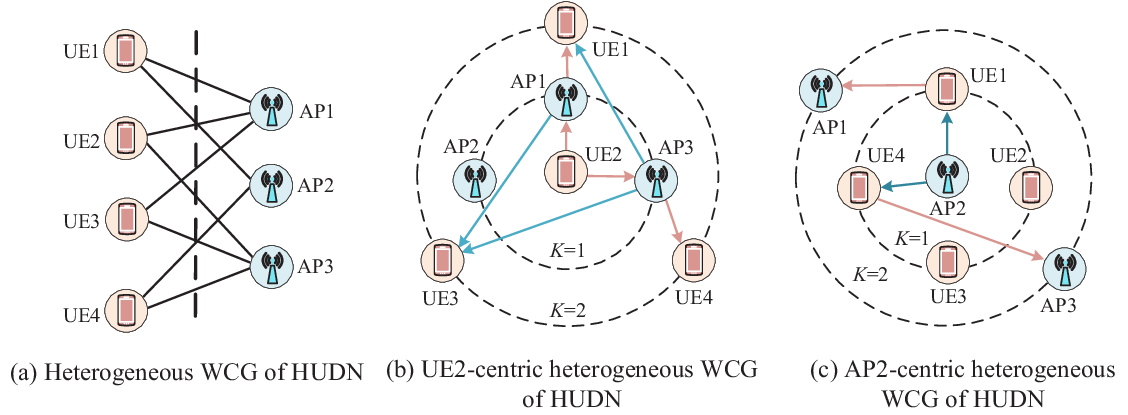} \\
	\caption{Illustration of constructing WCG for ultra-dense Cellular networks/WLANs~\cite{JUAPAinHUDN}.}
	\label{HGHGNN}
\end{figure}

In addition to the Cellular networks/WLANs scenarios aforementioned, there are also other scenarios, such as one AP may serve multiple UEs and one UE may access multiple APs, called Heterogeneous Ultra-Dense Network~(HUDN)~\cite{JUAPAinHUDN}. We treat the UEs and APs as two types of nodes, as shown in Fig.~\ref{HGHGNN}~(a). The feature vector of a node indicating UEs includes the CSIs to every AP. However, the feature vector of a node indicating APs includes the CSIs to every UE. The edge of WCG should be built only exists between nodes indicating UEs and nodes indicating APs when the UE can be detected. As shown in Fig.~\ref{HGHGNN}~(b), the first-order neighborhood, $K=1$, of UEs is the APs that have edge connect, the second-order neighborhood, $K=2$, of UEs is the UEs that connect to the first-order neighborhood. Similarly, the corresponding neighborhood of AP is shown in Fig.~\ref{HGHGNN}~(c). The details of designing the GNN model using this WCG can be found in~\cite{JUAPAinHUDN}, which exploit feature information from first-order and second-order neighborhood.

\section{Paradigms of GNNs}
\begin{figure}[t]
\renewcommand{\captionfont}{\footnotesize}
	\centering
\includegraphics[width=1.0\columnwidth,keepaspectratio]{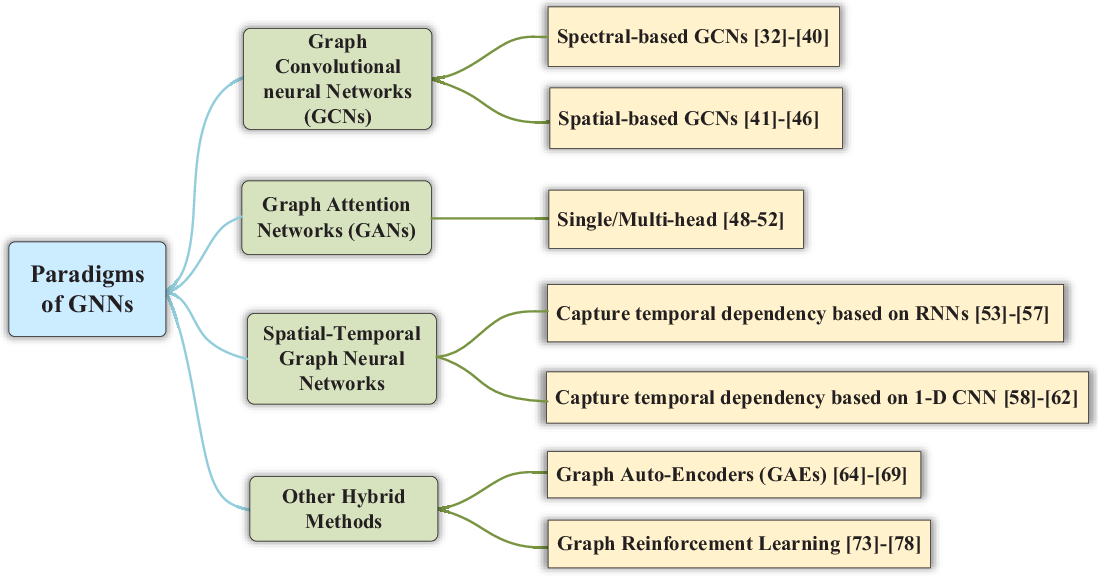}\\
	\caption{Overview of the paradigms of GNNs.}
	\label{paradigms}
\end{figure}

In this section, we simply introduce a few classical GNN models, mainly including graph convolutional neural networks, graph attention networks,  spatial-temporal graph neural networks, and other hybrid methods. The overview of these paradigms is shown in Fig.~\ref{paradigms}.

\subsection{Graph Convolutional Neural Networks}
Graph convolutional neural networks~(GCNs) implement the convolutional operation on graph-structured data, i.e., in non-Euclidean space~\cite{GeoDL2017}. The core idea of GCNs is to learn a mapping function, which can combine the neighbor nodes' information with its feature information to generate a new node representation. According to different convolution methods, GCNs can be divided into spectral-based GCNs~\cite{GCN,ChebNet,1stChebNet,DCN,AGC,Cayleyn,zhang2020spectrum,fu2020hesgcn,xu2021spectral} and spatial-based GCNs~\cite{SpatialGCN,MPNN2017,GraphSage2017,DCNN,pei2019geom,danel2020spatial}. In the sequel, we simply introduce several classical models of spectral-based GCNs and spatial-based GCNs, respectively.

\subsubsection{Spectral-based GCNs} Since the number of neighbors may be different for different nodes, a fixed convolutional kernel cannot be used on a graph. To address this problem, the graph-structured data is generally converted to the frequency domain. Specifically, for a given input graph signal $\mathbf{x}\in\mathbb{R}^d$ and a graph filter $\mathbf{g}\in\mathbb{R}^{d}$, the graph convolution is defined as~\cite{GNNsurvey12020}
\begin{equation}\label{Cachenable01}
\begin{aligned}
\mathbf{x}*_{G}\mathbf{g}&=\mathcal{F}^{-1}\left(\mathcal{F}(\mathbf{x})\odot \mathcal{F}(\mathbf{g})\right) \\
&=\mathbf{U}\left(\mathbf{U}^{T}\mathbf{x}\odot \mathbf{U}^{T}\mathbf{g}\right)=\mathbf{U}\hat{\mathbf{g}}\mathbf{U}^{T}\mathbf{x},
\end{aligned}
\end{equation}
where $*_{G}$ denotes the graph convolution operation, $\mathcal{F}(\mathbf{x})= \mathbf{U}^{T}\mathbf{x}$ denotes the graph Fourier transform, $\mathcal{F}^{-1}(\mathcal{F}(\mathbf{x}))= \mathbf{U}\mathcal{F}(\mathbf{x})$ denotes the inverse graph Fourier transform, $\odot$ denotes the Hadamard product, and  $\hat{\mathbf{g}}=\text{diag}\left(\mathbf{U}^T\mathbf{g}\right)$.

Various spectral-based GCNs have been defined by changing $\hat{\mathbf{g}}$. For example, Bruna \emph{et al.} proposed spectral CNN~(SpectralCNN) in which $\hat{\mathbf{g}}$ is learnable parameters~\cite{GCN}. However, due to the existing of the eigen-decomposition of $\tilde{\mathbf{L}}$, SpectralCNN faces several challenges, such as low computational efficiency of eigen-decomposition and the learned graph filters cannot be applied in a graph with different structure~\cite{GNNsurvey2019}. To overcome these shortcomings, Defferrard \emph{et al.} proposed Chebnet via redefining the graph filter with Chebyshev polynomials~\cite{ChebNet}. By constraining the number of parameters, Kipf \emph{et al.} further proposed a model named GCN, which has the ability to overcome the overfitting, to minimize the number of operations at each layer, i.e.,
\begin{equation}\label{Cachenable02}
\begin{aligned}
\mathbf{X}*_G=\mathbf{W}\left(\mathbf{I}_{N}+\mathbf{D}^{-\frac{1}{2}}\mathbf{AD}^{-\frac{1}{2}}\right)\mathbf{X},
\end{aligned}
\end{equation}
where $\mathbf{W}$ is a learnable weight matrix. In order to tackle the case in which  $\mathbf{I}_{N}+\mathbf{D}^{-\frac{1}{2}}\mathbf{AD}^{-\frac{1}{2}}\in(0,2)$ may lead to gradient explosion, the authors further transform $\mathbf{I}_{N}+\mathbf{D}^{-\frac{1}{2}}\mathbf{AD}^{-\frac{1}{2}}$ into $\widetilde{\mathbf{D}}^{-\frac{1}{2}}\widetilde{\mathbf{A}}\widetilde{\mathbf{D}}^{-\frac{1}{2}}$, where $\widetilde{\mathbf{A}}=\mathbf{A}+\mathbf{I}_{N}$ and $\widetilde{\mathbf{D}}_{ii}=\sum\nolimits_{j}\widetilde{\mathbf{A}}_{ij}$.
Compared to GCN, Chebnet has higher computational complexity, but it has stronger expression ability. Chebnet's $K$-order convolution operator can cover $K$ steps neighbor nodes of the central node, while GCN only covers the first-order neighbor nodes. However, the perception domain of graph convolution can be expanded by stacking multiple GCN layers, so the flexibility is relative high.

Y. Ma \emph{et al.} proposed a directed graph convolution network based on directed Laplacian, which is defined as~\cite{ma2019spectral}
\begin{equation}\label{Cachenable03}
\begin{aligned}
\mathbf{Z}=\frac{1}{2}\left(\tilde{\mathbf{\Phi}}^{\frac{1}{2}}\mathbf{P}\tilde{\mathbf{\Phi}}^{-\frac{1}{2}}+\tilde{\mathbf{\Phi}}^{-\frac{1}{2}}\tilde{\mathbf{P}}^T\tilde{\mathbf{\Phi}}^{\frac{1}{2}}\right)\mathbf{X}\mathbf{W},
\end{aligned}
\end{equation}
where $\tilde{\mathbf{D}}_{ii}^{(out)}=\sum_{j}\tilde{\mathbf{A}}_{i,j}$, $\tilde{\mathbf{P}}=\left(\tilde{\mathbf{D}}^{(out)}\right)^{-1}\tilde{\mathbf{A}}$ and $\tilde{\mathbf{\Phi}}$ is calculated based on $\tilde{\mathbf{P}}$. The directed graph filter is approximated by the first-order Chebyshev polynomials.

\begin{remark}
The application of spectral-based GCNs in wireless networks will be introduced in Section \uppercase\expandafter{\romannumeral4}. The definition of graph filters for spectral-based GCNs usually combines the adjacency matrix $\mathbf{A}$ of wireless network topology and the channel state information $\mathbf{H}$, which can make full use of the complex wireless information. On the other hand, $\mathbf{H}$ is expressed as the propagation relationship between nodes from the aspect of wireless network environment.
\end{remark}

\subsubsection{Spatial-based GCNs} The spatial-based graph convolution is similar to the image convolution. The two convolution operations all extract the neighbor information of a node to obtain a richer feature representation of the node or the pixel. The difference between image convolution and spatial-based graph convolution is that the nodes in a graph are unordered while the pixels in an image are irregular, and the number of neighbors of each pixel in an image is limited while the number of neighbors of each node in a graph is not sure. So, spatial-based graph convolution operation cannot use a fixed-size convolution kernel like the image convolution operation. Thus, the key of spatial-based GCNs is to define the convolution operation with different neighborhood numbers and keep local invariance.

The most widely used spatial-based GCNs in wireless networks are message passing neural network~(MPNN) and diffusion-convolutional neural networks~(DCNNs). MPNN was proposed in~\cite{MPNN2017}, which is a unified framework of spatial-based GCNs, and decomposes the spatial-based graph convolution into a message aggregation phase and a combination phase, i.e.,
\begin{subequations}\label{Cachenable04}
\begin{align}
\mathbf{m}_{v_i}^{(t)}&=\sum\limits_{v_{j}\in\mathcal{N}(v_i)}\mathcal{M}^{(t)}\left(\mathbf{X}_{i}^{(t-1)},\mathbf{X}_{j}^{(t-1)},\mathbf{e}_{ij}\right), \label{Cachenable04a}\\
\mathbf{X}_{i}^{(t)}&=\mathcal{U}^{(t)}\left(\mathbf{X}_{i}^{(t-1)},\mathbf{m}_{v_i}^{(t)}\right), \label{Cachenable04b}
\end{align}
\end{subequations}
where $\mathbf{e}_{ij}$ is the feature vector of the edge between nodes $v_i$ and $v_j$, $\mathcal{M}^{(t)}(\cdot)$ and $\mathcal{U}^{(t)}(\cdot)$ are the aggregation function and the combination function in the $t$-th iteration, respectively. $\mathbf{m}_{v_i}^{(t)}$ is the message aggregated from node $v_i$'s neighbors and $\mathbf{X}_{i}^{(t)}$ is the hidden state of node $v_i$ in the $t$-th iteration. It is observed that the computational efficiency of MPNN decreases with the increase of the number of nodes.

W. L. Hamilton further proposed Graph SAmple and aggreGatE~(GraphSAGE) model via fixing the number of neighbors for message passing to overcome the shortcomings of MPNN~\cite{GraphSage2017}. The graph convolution operation of GraphSAGE is implemented by
\begin{equation}\label{Cachenable05}
\begin{aligned}
\mathbf{X}_{v_i}^{(t)}=\sigma\left(\mathbf{W}^{(t)}g^{(t)}\left(\mathbf{X}_i^{(t-1)},\left\{\mathbf{X}_j^{(t-1)},\forall j\in S_{\mathcal{N}(v_i)}\right\}\right)\right),
\end{aligned}
\end{equation}
where $\mathbf{W}^{(t)}$ and $g^{(t)}(\cdot)$ are a learnable weight matrix and an aggregation function in the $t$-th layer, respectively. $\sigma\left(\cdot\right)$ is a nonlinear activation function. $S_{\mathcal{N}(v_i)}$ is a random sample of the node $v_i$'s neighbors. The main difference between MPNN and GraphSAGE is that GraphSAGE randomly samples a fixed number of neighbors for each node, while MPNN utilizes all the neighbors of each node. In addition, the diffusion-convolution operation in the DCNNs model builds a potential representation by scanning the diffusion process of each node through the transition probability matrix, i.e, $\mathbf{Z}^{(t)}=\sigma\left(\mathbf{W}^{(t)}\odot \mathbf{P}^{t}\mathbf{X}\right)$, where $\mathbf{Z}^{(t)}$ is the hidden state in the $t$-th layer and $\mathbf{P}^t$ denotes the $t$ power of $\mathbf{P}$.

\begin{remark}
The GCN models introduced above meet the graph structures with the same node type and edge type. However, in wireless networks, the types of communication devices and of communication mechanisms between devices may be diverse, that is, the corresponding WCG may be a heterogeneous. Therefore, the design of heterogeneous graph convolution is helpful to learn different types of information and is more suitable for the business needs of practical network scenarios. In Section \uppercase\expandafter{\romannumeral3}, several works on designing graph convolution for heterogeneous WCG are introduced. However, the existing methods are designed based on spatial-based GCNs. There are few works on designing heterogeneous graph convolution based on spectral-based GCNs.
\end{remark}

\subsection{Graph Attention Networks}
\begin{figure}[t]
\renewcommand{\captionfont}{\footnotesize}
	\centering
\includegraphics[width=1.0\columnwidth,keepaspectratio]{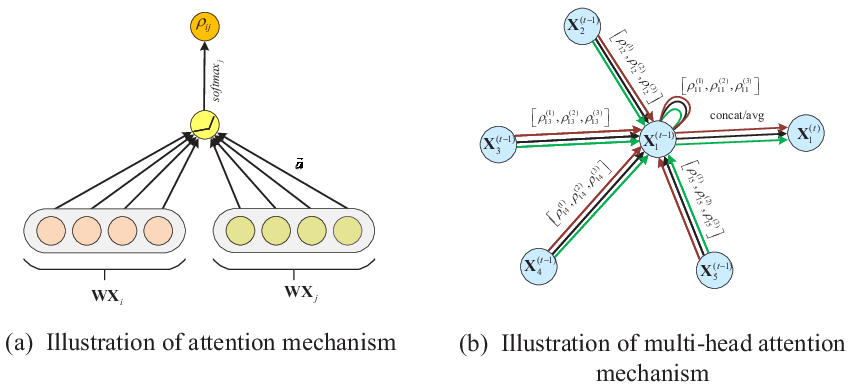}\\
	\caption{Illustration of the implementation details of attention mechanism~\cite{GAT}.}
	\label{GAT}
\end{figure}

One notes that in a graph, different neighbor nodes generally have different influences on the central node. This implies that one needs to distinguish the influences of nodes with a proper means during the design process of the learning model. Attention mechanism~\cite{Attention} has been regarded as an expressive means of information fusion by assigning weight to given information. In recent years, attention mechanism is also introduced into GNNs, such as, Velickovic \emph{et al}. proposed an attention mechanism based GNN model, i.e., Graph Attention Network~(GAT), by adaptively allocating weight to different neighbors in the aggregation operation~\cite{GAT}, which is defined as follows
\begin{equation}\label{Cachenable06}
\begin{aligned}
\mathbf{X}_{i}^{(t)}=\sigma\left(\sum_{j\in \mathcal{N}(v_i)}\alpha\left(\mathbf{X}_{i}^{(t-1)},\mathbf{X}_{j}^{(t-1)}\right)\mathbf{W}^{(t-1)}\mathbf{X}^{(t-1)}_{j}\right),
\end{aligned}
\end{equation}
where $\alpha(\cdot)$ is the attention mechanism, $\mathbf{X}^{(t)}_{i}$ is the hidden state of node $v_i$ at the $t$-th layer. In addition, multi-head attention mechanism is further introduced to improve the expression ability of the attention layer, that is, $K$ independent attention mechanisms can be utilized and then the output are concatenated together, i.e.,
\begin{equation}\label{Cachenable07}
\begin{aligned}
\mathbf{X}_{i}^{(t)}=\|^{K}_{k=1}\sigma\left(\sum_{j\in \mathcal{N}(v_i)}\alpha^{k}\left(\mathbf{X}_{i}^{(t-1)},\mathbf{X}_{j}^{(t-1)}\right)\mathbf{W}^{(t-1)}\mathbf{X}^{(t-1)}_{j}\right),
\end{aligned}
\end{equation}
where $\|$ denotes the concatenation operation and $\alpha^{k}(\cdot)$ is the $k$-th attention mechanism. The illustration of single-head attention mechanism and multi-head attention mechanism are shown in Fig.~\ref{GAT}, where $\rho_{ij}^{(k)}$ denotes the attention weight between node $v_i$ and node $v_j$ obtained by the $k$-th attention mechanism. Other GNNs models using attention mechanism could be found in~\cite{zhou2018commonsense,yang2019dynamic,li2019relation,wang2019heterogeneous}.

\begin{remark}
The essence of the attention mechanism is to weigh the information transmitted to distinguish the importance of different types of information. In wireless networks, especially heterogeneous wireless networks, the environmental information may be diverse, such as communication equipment, communication links, etc., which may affect the problems in different ways. Therefore, it is unreasonable to treat different types of information equally in the design of the learning model. On the contrary, one or more attention mechanisms should be designed according to the relevant prior knowledge to distinguish the influences of different types of information. There are few applications of graph attention mechanism in wireless networks.
\end{remark}

\subsection{Spatial-Temporal Graph Neural Networks}
Spatial-temporal GNNs~(STGNNs) play an important role in dealing with graphs that have dynamic node inputs while connected nodes are interdependent. There are two categories of STGNNs from the perspective of capturing temporal dependency, i.e., RNN-based methods~\cite{GraphGRUTP2019,geng2019spatiotemporal,yu2019st,guo2020optimized,li2017diffusion} and CNN-based methods~\cite{yu2017spatio,STGCN-tf2019,ge2019temporal,fang2019gstnet,wu2019graph}.

C. Chen \emph{et al.} utilized the residual recurrent GNN~(Res-RGNN) to predict the traffic flow in traffic network~\cite{GraphGRUTP2019}. Res-RGNN utilizes the spatial attributes to capture the spatial features with diffusion convolution, while using graph recurrent unit~(GRU) to discover the temporal dependency for each node. Specifically, the implementation of RGNN unit at time $t$ is
\begin{subequations}\label{Cachenable08}
\begin{align}
&\mathbf{r}^{(t)}=\sigma\left(\mathbf{\Theta}_r*_G\left[\mathbf{x}^{(t)},\mathbf{e}^{(t)},\mathbf{s}^{(t-1)}\right]+\mathbf{b}_r\right), \label{Cachenable8a} \\
&\mathbf{u}^{(t)}=\sigma\left(\mathbf{\Theta}_u*_G\left[\mathbf{x}^{(t)},\mathbf{e}^{(t)},\mathbf{s}^{(t-1)}\right]+\mathbf{b}_u\right), \label{Cachenable8b} \\
&\mathbf{c}^{(t)}=\mathrm{tanh}\left(\mathbf{\Theta}_c*_G\left[\mathbf{x}^{(t)},\mathbf{e}^{(t)},(\mathbf{r}^{(t)}\odot \mathbf{s}^{(t-1)})\right]+\mathbf{b}_c\right), \label{Cachenable8c}
\\
&\mathbf{s}^{(t)}=\mathbf{u}^{(t)}\odot \mathbf{s}^{(t-1)}+\left(1-\mathbf{u}^{(t)}\right)\odot \mathbf{c}^{(t)}, \label{Cachenable8d}\\
&\mathbf{y}^{(t+1)}=\mathbf{W}_o\mathbf{s}^{(t)}, \label{Cachenable8e}
\end{align}
\end{subequations}
where $\mathbf{x}^{(t)},\mathbf{e}^{(t)}$ and $\mathbf{s}^{(t)}$ denote the graph signal, external feature and the outputted hidden state at time $t$, respectively. $\mathbf{r}^{(t)}$ and $\mathbf{u}^{(t)}$ represent the reset gate and update gate at time $t$, respectively. $\mathbf{\Theta}_r,\mathbf{\Theta}_u$ and $\mathbf{\Theta}_c$ are the learnable graph filters, $\mathbf{W}_o$ is the learned weights of the output layer. $\mathbf{y}^{(t+1)}$ denotes the output at time $t+1$.

\begin{figure}[t]
\renewcommand{\captionfont}{\footnotesize}
	\centering
\includegraphics[width=1.0\columnwidth,keepaspectratio]{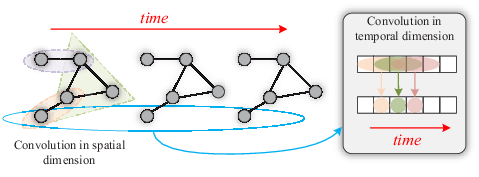}\\
	\caption{Illustration of how ASTGCN captures spatial and temporal dependencies.}
	\label{ASTGCN}
\end{figure}

S. Guo \emph{et. al} proposed an attention mechanism based spatial-temporal GNN~(ASTGCN) to predict traffic flow in traffic network~\cite{STGCN-tf2019}. Specifically, ASTGCN utilizes the spectral-based GCNs, i.e., ChebNet~\cite{ChebNet}, to capture the spatial dependency among different nodes in traffic network graph. Meanwhile, one dimension CNN is utilized to capture temporal dependency for each node in time series. The implementation details of capturing spatial and temporal dependencies are illustrated in Fig.~\ref{ASTGCN}.

\begin{remark}
With the development of communication technologies, the wireless network becomes more complex and huge with massive terminals. As a result, the resource management of wireless networks becomes more and more challenging. To improve spectral efficiency, prediction plays an important role in wireless networks. Applying STGNNs to traffic prediction has attracted extensive attention in both academic and industry, which contributes to the resource management of wireless networks. Of course, other directions not involved in this overview also need to consider the spatial-temporal dependencies during designing the GNNs for wireless networks.
\end{remark}

\subsection{Other Hybrid Methods}
\begin{figure}[t]
\renewcommand{\captionfont}{\footnotesize}
	\centering
\includegraphics[width=1.0\columnwidth,keepaspectratio]{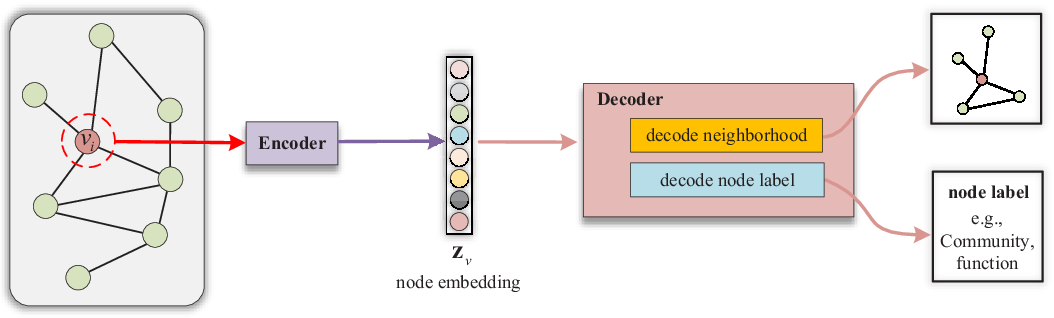}\\
	\caption{Overview of GAEs.}
	\label{GAE}
\end{figure}

Inspired by the conventional auto-encoders, graph auto-encoders~(GAEs) utilizing GNNs as encoders to learn low-dimensional latent representations~(or embeddings) of nodes have been investigated for wireless networks. The goal of encoders in GAEs is to encode the structural information of nodes. While the decoder in GAEs aims at decoding the structural information about the graph from the learned latent representations~\cite{RepresentLearnGraph2017}. The general overview of GAEs is shown in Fig.~\ref{GAE}. Specifically, the encoder maps node $v$ to a low-dimensional embedding vector $\mathbf{z}_v$ based on the node's structural information, and the decoder extracts the information interested from the low-dimensional embedding vector. GAEs have been used in many fields by virtue of their concise encoder-decoder structure and efficient encoding ability~\cite{VGAE,ARGA,choong2019optimizing,do2020graph,fan2020one2multi,li2020r}. Kipf \emph{et al.} proposed the variational GAE using a GCN encoder and a simple inner product decoder, which aims at the link prediction in citation networks~\cite{VGAE}. The encoder maps each node to a low-dimensional latent representation using GCN, then a network embedding $\mathbf{Z}$ can be obtained. The decoder computes the pair-wise distance given network embedding and applies a non-linear activation. Finally, the decoder outputs the reconstructed adjacency matrix.

In recent years, reinforcement learning~(RL) has been gradually applied to graph-structured tasks, such as graph generation~\cite{svetlik2017automatic,GCPN2018}, graph classification~\cite{GAM2018}, and graph reasoning tasks~\cite{DeepPath2017,wang2020adrl,wan2020reasoning,tiwari2021dapath,li2021memorypath,chen2021rlpath}, etc. J. You \emph{et al.} proposed a graph convolutional policy network~(GCPN) based on RL and GCNs to address the problem of non-differentiable objective functions and constraints~\cite{GCPN2018}. Graph attention model~(GAM) was proposed based on RL and random walks to solve the graph classification task~\cite{GAM2018}. The generation of random walks was modeled as a partially observable Markov decision process. The RL agent performs two actions at each time step, i.e., predicts the label of input graph and generates the rank vector using designed rank network. The reward is designed as $\mathcal{J}(\theta)=\mathbb{E}_{P(S_{1:L};\theta)}\sum_{l=1}^{L}r_l$, where $r_l=1$ if the GAM classified the graph correctly, otherwise, $r_l=-1$. $S_l$ is the environment. W. Xiong \emph{et al.} proposed a DeepPath model to find the most informative path between two target nodes with the goal of solving the knowledge graph reasoning task~\cite{DeepPath2017}. The action of RL agents is to predict the next node in the path at each step and output a reasoning path in the knowledge graph. The reward functions include the scoring criteria: global accuracy, path efficiency and path diversity.

\begin{remark}
The main advantage of GAEs is to mine the topological information in the graph, and then learn an effective low dimensional feature vector representation for each node or the whole graph. This feature vector representation can reflect the characteristic that can separate from other nodes or graphs to a certain extent. Although RL has been widely used in wireless networks, the application of RL in GNNs is still in its infancy. Generally speaking, the introduction of RL can enable GNNs to achieve approximate optimal performance without the prior information of the environment, and have independent exploration and optimal decision-making capabilities. Therefore, the introduction of RL into wireless communication technology has important practical significance.
\end{remark}

\section{Applications in Wireless Networks}
\begin{figure*}[t] \tiny
\renewcommand{\captionfont}{\footnotesize}
	\centering
\includegraphics[width=1.5\columnwidth,keepaspectratio]{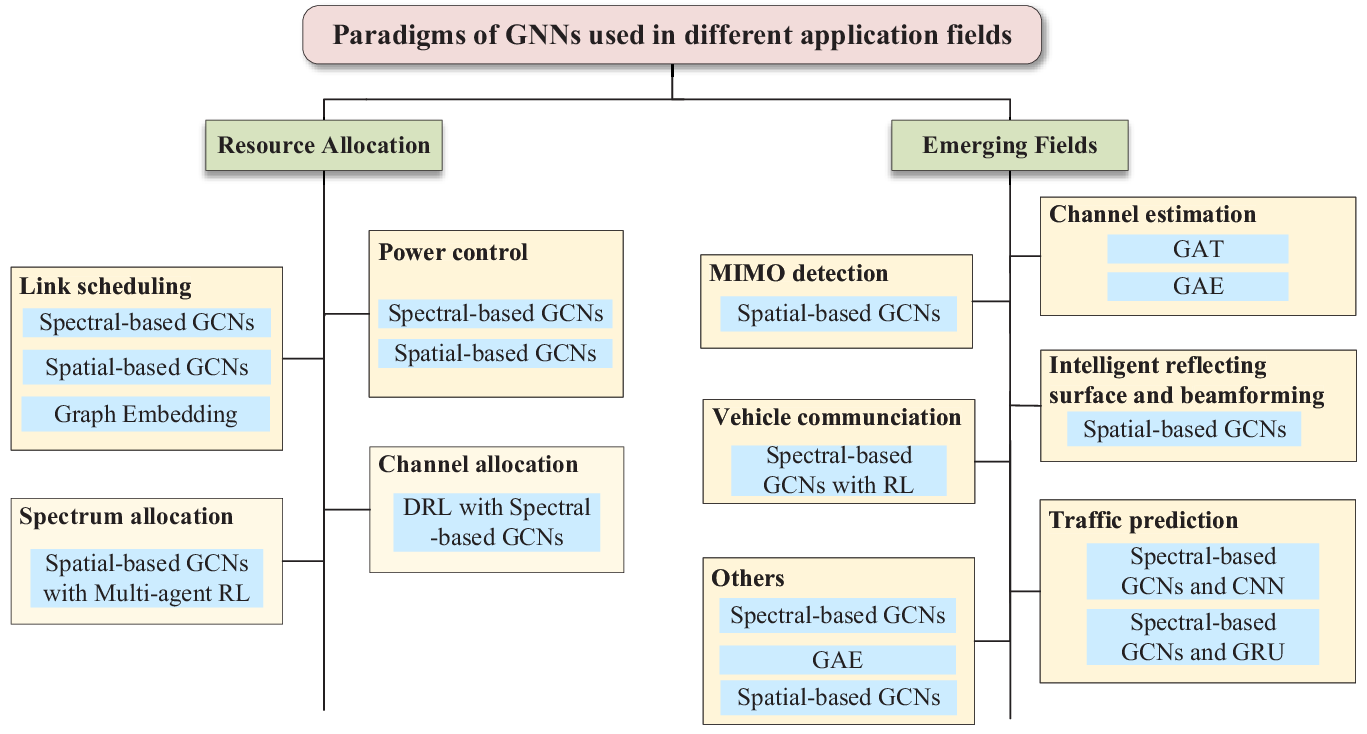}\\
	\caption{Paradigms of GNNs used in different application fields.}
	\label{TechUsed}
\end{figure*}

In this section, we focus on introducing comprehensively the application of GNNs in wireless networks. As shown in Table \uppercase\expandafter{\romannumeral2}, the applications of GNNs in wireless networks mainly cover resource allocation and a few emerging fields. The commonly used algorithms are illustrated in Fig.~\ref{TechUsed} for each research direction in wireless networks.
\begin{table*}[h!t]
  \centering
  \caption{\textsc{Applications of GNNs in Wireless Networks}}
\begin{tabular}{c l l l l l} \hline\hline
 \makecell[l]{\textbf{Area}} & \makecell[l]{\textbf{Year}} & \makecell[l]{\textbf{Application}} & \makecell[c]{\textbf{Wireless} \\\textbf{network type}} & \makecell[l]{\textbf{Algorithm}} & \makecell[l]{\textbf{Scheme}} \\ \hline
 \multirow{22}{*}{\makecell[l]{Resource \\Allocation}} & 2019 & \makecell[l]{Power control in ad-hoc wireless networks} & \makecell[l]{Mesh/Ad-hoc} & Spatial-based GCNs &  Y. Shen~\emph{et al.}~\cite{gnnWPC2019} \\
 \cline{2-6}
  & 2021 & \makecell[l]{Radio resource management in mmWave networks} & \makecell[l]{Mesh/Ad-hoc} & Spatial-based GCNs & Y. Shen~\emph{et al.}~\cite{gnnResourceMng2021} \\
 \cline{2-6}
  & 2019 & \makecell[l]{Power allocation with a set
of transmitter \\receiver pairs in a large scale wireless network} & \makecell[l]{Mesh/Ad-hoc} & Spectral-based GCNs & M. Eisen~\emph{et al.}~\cite{LSWgnn2019}\\
 \cline{2-6}
 & 2020 & Power allocation in wireless networks & \makecell[l]{Ad-hoc/Cellular} & Spectral-based GCNs &  M. Eisen~\emph{et al.}~\cite{REGNN2020}\\
 \cline{2-6}
  & 2020 & \makecell[l]{Power control in ad-hoc wireless networks} & \makecell[l]{Mesh/Ad-hoc} & Spectral-based GCNs & M. Eisen~\emph{et al.}~\cite{TPgnn2020}\\
 \cline{2-6}
  & 2021 & \makecell[l]{Power control in decentralized wireless networks} & \makecell[l]{Mesh/Ad-hoc} & Spectral-based GCNs &  I. Nikoloska~\emph{et al.}~\cite{FastPowerControl2021} \\
 \cline{2-6}
  & 2020 & Downlink power control
in wireless networks & \makecell[l]{Mesh/Ad-hoc} & Spectral-based GCNs & N. Naderializadeh~\emph{et al.}~\cite{wpcCOgnn2020}\\
 \cline{2-6}
 & 2021 & \makecell[l]{Power control/beamforming in heterogeneous\\ D2D networks} & \makecell[l]{Mesh/Ad-hoc} & Spatial-based GCNs &  X. Zhang~\emph{et al.}~\cite{PowerContrlBF2021} \\
 \cline{2-6}
 & 2021 & Power control in multi-cell cellular networks & Cellular & Spatial-based GCNs &  J. Guo~\emph{et al.}~\cite{LpcCSgnn} \\
 \cline{2-6}
 & 2021 & \makecell[l]{Joint user association and power allocation in \\
heterogeneous ultra dense network} & \makecell[l]{Cellular} & Spatial-based GCNs & X. Zhang~\emph{et al.}~\cite{JUAPAinHUDN} \\
 \cline{2-6}
 & 2020 & \makecell[l]{Resource allocation
in free space optical~(FSO)\\ fronthaul networks} & \makecell[l]{Mesh/Ad-hoc} & Spectral-based GCNs &  Z. Gao~\emph{et al.}~\cite{ResAllocGNN} \\
 \cline{2-6}
 & 2020 & \makecell[l]{Resource allocation problems under \\asynchronous wireless network setting} & \makecell[l]{Mesh/Ad-hoc} & Spectral-based GCNs & Z. Wang~\emph{et al.}~\cite{UspLrnARAadhoc, wang2021learning} \\
 \cline{2-6}
  & 2021 & \makecell[l]{Resource allocation in wireless IoT networks} & \makecell[l]{Mesh/Ad-hoc} & Spatial-based GCNs & T. Chen~\emph{et al.}~\cite{chen2021gnn} \\
 \cline{2-6}
  & 2020 & \makecell[l]{Power allocation in a single-hop
ad-hoc \\wireless network} & \makecell[l]{Mesh/Ad-hoc} & Spectral-based GCNs & A. Chowdhury~\emph{et al.}~\cite{FpaGNNdau,UnfoldGnnPA}\\
 \cline{2-6}
 & 2019 & \makecell[l]{Link scheduling in D2D networks} & \makecell[l]{Mesh/Ad-hoc} & Graph embedding & W. M. Lee~\emph{et al.}~\cite{graphEmbedWLS} \\
 \cline{2-6}
  & 2020 & \makecell[l]{Schedule transmission for wireless networks \\in a distributed manner} & \makecell[l]{Mesh/Ad-hoc} & Spectral-based GCNs & Z. Zhao~\emph{et al.}~\cite{DistributeSchedgnn} \\
 \cline{2-6}
  & 2019 & \makecell[l]{Temporal
link prediction in various network systems} & \makecell[l]{Mesh/Ad-hoc} & Spectral-based GCNs & K. Lei~\emph{et al.}~\cite{Gcn-gcnLP2019}\\
 \cline{2-6}
 & 2021 & \makecell[l]{Joint link scheduling and beam selection in \\ultra-dense D2D mmWave communication networks} & \makecell[l]{Mesh/Ad-hoc} & Spatial-based GCNs & S. He~\emph{et al.}~\cite{he2021gblinks}\\
 \cline{2-6}
 & 2020 & \makecell[l]{Channel allocation for densely deployed WLANs} & WLANs & \makecell[l]{DRL with Spectral-\\based GCN} & K. Nakashima~\emph{et al.}~\cite{DeepRLCAgnn2020} \\
 \cline{2-6}
 & 2020 & \makecell[l]{Spectrum allocation in vehicle-to-everything~(V2X) \\networks} & \makecell[l]{Mesh/Ad-hoc} & \makecell[l]{Spatial-based GCNs \\ with Multi-agent RL} &  Z. He~\emph{et al.}~\cite{RAgnnVC2020}\\
 \cline{2-6}
 & 2021 & \makecell[l]{AP selection for Cell-Free massive MIMO systems} & \makecell[l]{Celluar} & \makecell[l]{Spatial-based GCNs} &  V. Ranasinghe~\emph{et al.}~\cite{ranasinghe2021graph}\\
 \hline
 \multirow{16}{*}{\makecell{Emerging \\ Fields}} & 2021 & \makecell[l]{Intelligent reflecting surface and beamforming} & Celluar & Spatial-based GCNs & T. Jiang~\emph{et al.}~\cite{BeaformIRS2021}\\
 \cline{2-6}
  & 2020 & \makecell[l]{Channel estimation for wireless networks} & Celluar & GAT & K. Tekb ıy ık~\emph{et al.}~\cite{ChannelEstGAN}\\
 \cline{2-6}
  & 2020 & \makecell[l]{Massive MIMO detection in wireless communication} & \makecell[l]{Mesh/Ad-hoc} & Spatial-based GCNs & A. Scotti~\emph{et al.}~\cite{GnnMIMOdetect}\\
 \cline{2-6}
 & 2020 & \makecell[l]{Channel tracking for the
massive MIMO networks} & Celluar & GAE &  Y. Yan~\emph{et al.}~\cite{GnnCTMIMO2020} \\
 \cline{2-6}
 & 2020 & \makecell[l]{Cellular traffic prediction} & Celluar & \makecell[c]{Spectral-based GCNs \\ and CNN with GLU} & S. Zhao~\emph{et al.}~\cite{CellNetTPgnn2020} \\
 \cline{2-6}
  & 2020 & \makecell[l]{Satellite traffic prediction} & \makecell[l]{Mesh/Ad-hoc} & \makecell[l]{Spectral-based GCNs \\with graph GRU} & L. Yang~\emph{et al.}~\cite{StlNetTPgcngru2020} \\
 \cline{2-6}
 & 2020 & \makecell[l]{Multiagent cooperative control for CAV networks} & \makecell[l]{Mesh/Ad-hoc} & \makecell[l]{Spectral-based GCNs \\ with RL} &  J. Dong~\emph{et al.}~\cite{GCQNcavNet} \\
 \cline{2-6}
 & 2020 & \makecell[l]{Active traffic management for CAV networks} & \makecell[l]{Mesh/Ad-hoc} & \makecell[l]{Spectral-based GCNs \\ with RL} &  P. Y. J. Ha~\emph{et al.}~\cite{cavMARL} \\
 \cline{2-6}
 & 2020 & \makecell[l]{Efficient point cloud processing} & \makecell[l]{Mesh/Ad-hoc} & Dynamic GCNs &  J. Shao~\emph{et al.}~\cite{Branchy-GNN} \\
 \cline{2-6}
 & 2020 & \makecell[l]{Point cloud delivery} & \makecell[l]{Mesh/Ad-hoc} & GAE & T. Fujihashi~\emph{et al.}~\cite{Wireless3Dpcgnn}\\
 \cline{2-6}
 & 2021 & \makecell[l]{3D object detection} & \makecell[l]{Mesh/Ad-hoc} & 3D GNN & C. S. Jeong~\emph{et al.}~\cite{ARanchor2021}\\
 \cline{2-6}
 & 2019 & \makecell[l]{Throughput maximization for UAV assisted \\ground networks} & \makecell[l]{Mesh/Ad-hoc} & Spectral-based GCNs &  S. Lohani~\emph{et al.}~\cite{FSOAMC2019} \\
 \cline{2-6}
 & 2021 & \makecell[l]{Wireless network localization} & - & Spectral-based GCNs & W. Yan~\emph{et al.}~\cite{NetworkLoc2021}\\
 \cline{2-6}
 & 2021 & \makecell[l]{Performance prediction in Next-Generation WLANs} & WLANs & Spatial-based GCNs & P. Soto~\emph{et al.}~\cite{soto2021atari}\\
 \cline{2-6}
 & 2021 & \makecell[l]{Routing in small satellite networks} & \makecell[l]{Mesh/Ad-hoc} & Spectral-based GCNs & M. Liu~\emph{et al.}~\cite{liu2021routing}\\
 \cline{2-6}
 & 2021 & \makecell[l]{Decentralized control in wireless \\communication systems} & \makecell[l]{Mesh/Ad-hoc} & Spectral-based GCNs & M. Lee~\emph{et al.}~\cite{DecentralizedInfer2021}\\
 \hline\hline
\end{tabular}
\end{table*}

\subsection{Resource Allocation}
Resource allocation is one of the key issues for wireless communication systems. Applying GNNs to study the problem of resource allocation mainly focuses on power control, link scheduling, channel allocation, and spectrum allocation, etc.

\setcounter{subsubsection}{0}
\subsubsection{Power control} A large amount of works has studied the power control problem using traditional optimization methods and DNNs. Unfortunately, the traditional optimization methods face high computational complexity. On the other hand, with the expansion of the wireless network scale, the scalability and generalization of DNNs will become worse. Motivated by these observations, many researchers utilize the GNNs, which have the natural characteristics of solving the problem with graph-structured data, to investigate the power control problem in wireless networks.

To develop scalable methods to solve the power control problem in wireless networks, Y. Shen \emph{et al.} proposed an interference graph convolutional neural network~(IGCNet) based on MPNN for $K$-user interference channels~\cite{gnnWPC2019}. In particular, the $K$-user interference channels are modeled as a complete WCG with node and edge labels, as shown in Fig.~\ref{K-user-interference}. The aggregation and combination rules of IGCNet are designed as follows
\begin{subequations}\label{Cachenable09}
\begin{align}
\bm{\gamma}_{j,i}^{(t)}&=\mathrm{MLP1}\left(h_{ji},h_{ij},w_i,h_{jj},\bm{\beta}_j^{(t-1)}\right), &\\
\bm{\alpha}_i^{(t)}&=\mathrm{CONCAT}\left(\mathrm{MAX}_{j\in\mathcal{N}(v_i)}\left(\bm{\gamma}_{j,i}^{(t)}\right),\sum_{j\in\mathcal{N}(v_i)}\bm{\gamma}_{j,i}^{(t)}\right), &\\
\bm{\beta}_i^{(t)}&=\mathrm{MLP2}\left(\bm{\alpha}_i^{(t)},h_{ii},\bm{\beta}_i^{(t-1)},w_i\right), &
\end{align}
\end{subequations}
where $\mathrm{MAX\left({\cdot}\right)}$ is to take the largest value in a set, $\mathrm{MLP1}$ and $\mathrm{MLP2}$ represent two different $\mathrm{MLPs}$, $\mathrm{CONCAT}$ denotes the operation of vector concatenations. $\bm{\gamma}_{j,i}^{(t)}$ denotes the feature vector of the edge connecting node $v_j$ and node $v_i$ in the $t$-th iteration. $\bm{\alpha}_i^{(t)}$ is the aggregated information from the neighbor nodes to the central node $v_i$, and $\bm{\beta}_i^{(t)}$ is the updated hidden representation of node $v_i$ in the $t$-th iteration. The IGCNet is trained in an unsupervised manner to learn the optimal power control.

In addition, a family of neural networks, i.e., message passing graph neural networks~(MPGNNs), is designed to solve the problem of radio resource management in wireless networks~\cite{gnnResourceMng2021}. It demonstrates that MPGNNs satisfy the permutation equivariance property and have the ability to address the resource management problem of large-scale wireless networks while enjoying a high computational efficiency. To guarantee an effective implementation, this work further proposed a wireless channel graph convolution network~(WCGCN) belonging to the MPGNNs class. The effectiveness of WCGCN is evaluated with respect to the power control and beamforming problems. It demonstrates that WCGCN matches or outperforms the classic optimization-based algorithms and does not need domain knowledge and has significant speedups. However, MPGNNs just consider the problem with simple constraints or without constraints, complex resource constraints need to be further considered.

To solve the power allocation problem for device-to-device~(D2D) wireless networks, Spectral-based GCNs was employed in~\cite{LSWgnn2019,REGNN2020,TPgnn2020}. The proposed model, i.e., the random edge graph neural networks~(REGNN), performs the convolutions over a random graph formed by the fading interference patterns in wireless networks. The authors further presented an unsupervised model-free primal-dual learning algorithm to train the weights of the REGNN to overcome the difficulties incurred by the constrained objective function. Additionally, REGNN is utilized to solve the problem of power control in decentralized wireless networks~\cite{FastPowerControl2021}. To adapt the time-varying topologies, the first-order meta-learning is adopted to adapt the new network configurations with a few shots exploiting the data obtained from multiple topologies. The problem of downlink power control in wireless networks over a single shared wireless medium is investigated and addressed by using spectral-based GCNs and primal-dual learning~\cite{wpcCOgnn2020}. The main highlight of the works aforementioned is to solve the resource management problem under the complex constraints via the primal-dual learning method in homogeneous wireless networks.

Compared with the homogeneous wireless networks, it is more challenging to design the GNNs-based learning mechanism for the resource allocation problem in heterogeneous wireless networks. X. Zhang \emph{et al.} focused on addressing the problem of power control or beamforming using MPNN in heterogeneous D2D networks~\cite{PowerContrlBF2021}. This work considers a heterogeneous D2D network with two types of links, in which each kind of link holds different features, as depicted in Fig.~\ref{heterograph}. In particular, let $r=(n,m)$ be the interference from link type $n$ to link type $m$, the update rules in relation $(n,m)$ is defined as follows
\begin{subequations}\label{Cachenable10}
\begin{align}
&\mathbf{e}_{j_ni_m}[l]=\phi_{(n,m)}^e\left(\mathbf{v}_{j_n}[l-1],\mathbf{e}_{j_ni_m}[0]\right), \\
&\mathbf{v}_{i_m}^{(n)}=\phi_{(n,m)}^v\left(\mathbf{v}_{i_m}[l-1], \max\limits_{j\in{\mathcal{N}_{i_m}^{(n)}}}\mathbf{e}_{j_ni_m}[l]\right).
\end{align}
\end{subequations}
where $\phi_r^e$ and $\phi_r^v$ are an edge update function and a node update function of relation $r$, respectively. The aggregation rules is given by
\begin{equation}\label{Cachenable11}
\begin{aligned}
\mathbf{v}_{i_m}[l]=\rho_m^{v\rightarrow{v}}\left(\left\{\mathbf{v}_{i_m}^{(n)}[l]\right\}_n\right)=\frac{1}{c_{i,m}}\sum\limits_{n}\mathbf{v}_{i_m}^{(n)}[l],
\end{aligned}
\end{equation}
where $c_{i,m}$ is the number of relations causing interference to link $i_m$. $\rho_m^{v\rightarrow{v}}\left(\cdot\right)$ is the aggregation function of node to node with link type $m$.

Similarly, J. Guo~\emph{et al.} considered the power control problem in multi-cell cellular networks~\cite{LpcCSgnn}. Specifically, this work models the cellular networks as a heterogeneous graph, i.e., wireless interference graph, and then proposed a heterogeneous GNN~(HetGNN) based on spatial-based GCNs, called PGNN, to learn the power control policy in multi-cell cellular networks. Inspired by the finding that the parameter sharing scheme determines the invariance or equivalence relationship, the optimal power control policy has a combination of different PI and PE properties that existing heterogeneous GNNs do not satisfy~\cite{EqParaShare}. Additionally, X. Zhang~\emph{et al.} considered the joint user association and power control problem in HUDNs~\cite{JUAPAinHUDN}. The HUDNs are also modeled as a heterogeneous graph, which is shown in Fig.~\ref{HGHGNN}. A heterogeneous GraphSAGE~(HGSAGE) that extended from GraphSAGE~\cite{GraphSage2017}, is used to extract the latent node representations. To embrace both the generalization of the learning algorithm and the higher performance of HUDNs, the learning process of HUDNs is divided into two phases. The first phase of HUDNs learns a representation with a tremendous generalized ability to suit any scenario with different user distributions in an off-line manner. The second phase of HUDNs is to finely tune the parameters of GNN online to further improve the performance for quasi-static user distribution.

Z. Gao~\emph{et al.} investigated the optimal power assignment and node selection based on the instantaneous channel state information of the links in free space optical~(FSO) fronthaul networks~\cite{ResAllocGNN}. Spectral-based GCNs are utilized to exploit the FSO network structure with small-scale training parameters. Then, a primal-dual learning algorithm is developed to train the GNN in a model-free manner. Z. Wang~\emph{et al.} addressed the asynchronous decentralized wireless resource allocation problem with a novel unsupervised learning approach~\cite{UspLrnARAadhoc, wang2021learning}. Specifically, the interference patterns between transmitting devices are modeled as a graph to capture the asynchrony patterns via the activation of the edges on a highly granular time scale. A decentralized learning architecture, i.e., the aggregation graph neural networks~(Agg-GNNs) is designed based on the graph representation of interference and asynchrony. T. Chen $et~al.$ proposed a spatial-based GCNs based framework to address the high complexity of the practical implementation of wireless internet of things~(IoT) networks~\cite{chen2021gnn}. The effectiveness of the framework is evaluated by the link scheduling in D2D networks and the joint channel and power allocation in D2D underlaid cellular networks.

\begin{figure}[t]
\renewcommand{\captionfont}{\footnotesize}
	\centering
\includegraphics[width=1.0\columnwidth,keepaspectratio]{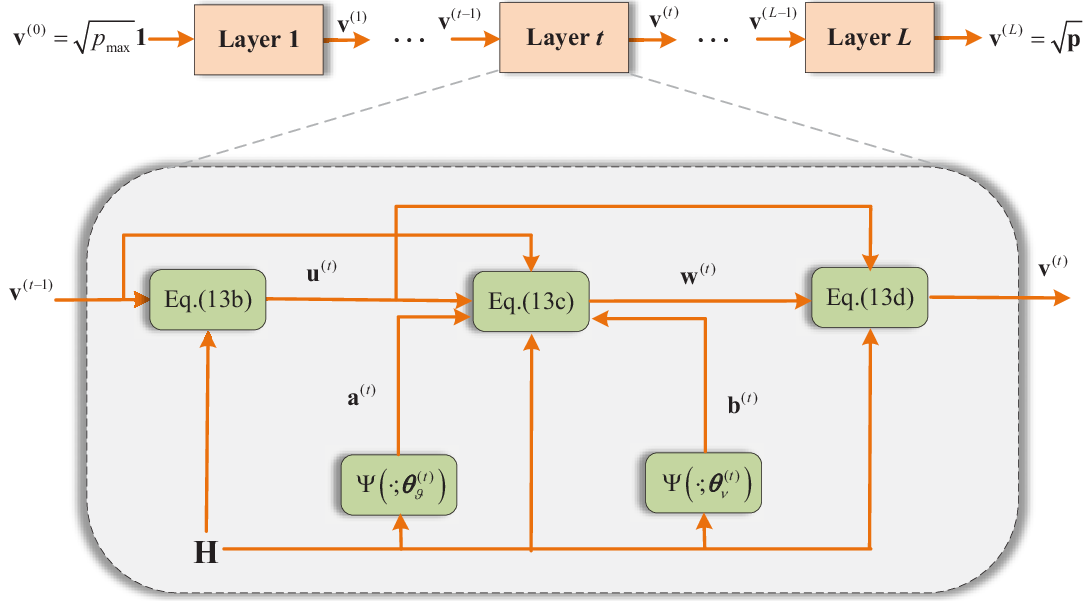}\\
	\caption{Illustration of UWMMSE~\cite{FpaGNNdau}.}
	\label{unfoldWMMSE}
\end{figure}

The methods proposed by the aforementioned works all are based on data-driven neural networks with poor interpretability and scalability. Inspired by the algorithmic unfolding of the iterative WMMSE, i.e., unfolded WMMSE~(UWMMSE), A. Chowdhury \emph{et al.} proposed a data- and model-driven neural architecture to solve the power allocation problem in a single-hop Ad-hoc wireless network~\cite{FpaGNNdau}. The optimization problem that should be solved is
\begin{subequations}\label{Cachenable12}
\begin{align}
& \ \min\limits_{\mathbf{w},\mathbf{a},\mathbf{b}}~\sum\limits_{i=1}^{M}\left(w_iq_i-\mathrm{log}w_i\right), \\
s.t.~ &\ q_i=\left(1-a_ih_{ii}b_i\right)^2+\sigma^2a_i^{2}+\sum\limits_{i\neq j}a_i^{2}h_{ij}^{2}b_j^{2}, \\
& \ b_i^{2}\leq p_{max},
\end{align}
\end{subequations}
where $\mathbf{w}=\left[w_1,w_2,...,w_M\right]^T,\mathbf{a}=\left[a_1,a_2,...,a_M\right]^T, \mathbf{b}=\left[b_1,b_2,...,b_M\right]^T$ are vectors of optimization variables. The allocated power is computed by a function $\mathbf{p}=\mathrm{\Phi}\left(\mathbf{H};\bm{\theta}_{\vartheta}, \bm{\theta}_{\nu}\right)$ of the channel state matrix through a layered architecture $\Phi$ with trainable weights $\bm{\theta}_{\vartheta}$ and $\bm{\theta}_{\nu}$. Precisely, setting $\mathbf{b}^{(0)}=\sqrt{p_{max}}\mathbf{1}$, the $t$-th layer of UWMMSE is implemented as follows
\begin{subequations}\label{Cachenable13}
\begin{align}
&\mathbf{\vartheta}^{(t)}=\Psi\left(\mathbf{H};\bm{\theta}_{\vartheta}^{(t)}\right),~~\mathbf{\nu}^{(t)}=\Psi\left(\mathbf{H};\bm{\theta}_{\nu}^{(t)}\right), \\
&a_i^{(t)}=\frac{h_{ii}{\nu}_i^{(t-1)}}{\sigma^2+\sum\limits_{j}h_{ij}^2b_j^{(t-1)}b_j^{(t-1)}}, \\
&w_i^{(t)}=\frac{{\vartheta}_i^{(t)}}{1-a_i^{(t)}h_{ii}b_i^{(t-1)}}+{\nu}_i^{(t)}, \\
&b_i^{(t)}=\alpha\left(\frac{a_i^{(t)}h_{ii}w_i^{(t)}}{\sum\limits_{j}h_{ji}^2a_{j}^{(t)}a_{j}^{(t)}w_{j}^{(t)}}\right),
\end{align}
\end{subequations}
and the output power is determined as $\mathbf{p}=\Phi\left(\mathbf{H};\bm{\theta}_{\vartheta}, \bm{\theta}_{\nu}\right)=\left(\mathbf{b}^{(L)}\right)^2$. $\alpha(z):=[z]_0^{\sqrt{p_{max}}}$ simply ensures that $b_i^{(t)}\in{\left[0,\sqrt{p_{max}}\right]}$. The function $\Psi$ parameterized by $\bm{\theta}_{\vartheta}$ and $\bm{\theta}_{\nu}$ is chosen to be spectral-based GCNs. The whole workflow of UWMMSE is shown in Fig.~\ref{unfoldWMMSE}, which has better interpretability and scalability compared with the data-driven learning models. Numerical experiments demonstrate that UWMMSE not noly significantly reduced the computational complexity, but also improved the performance compared to the conventional WMMSE~\cite{UnfoldGnnPA}.

\subsubsection{Link scheduling} Although the overall performance of wireless networks can be improved via power control, it is not enough to eliminate the strong interference for ultra-dense wireless networks. Link scheduling is regarded as an effective means to further improve the performance of wireless networks.

To overcome the high computational complexity of the traditional optimization methods and eliminate the costly channel estimation, M. Lee~\emph{et al.} proposed a novel DL-based graph embedding method to implement the link scheduling in D2D networks~\cite{graphEmbedWLS}. In detail, this work firstly models the D2D network as a fully connected directed graph, then computes a low-dimensional feature vector based on the distances of both communication and interference links without requiring the accurate channel state information for each node. Finally, a multi-layer classifier is utilized to learn the scheduling policy in a supervised and unsupervised manner, respectively. Numerical results show that this method can achieve near-optimal performance compared with the state-of-the-art methods but with a small number of training samples, and has competitive generalization and scalability.

A distributed scheduling scheme was proposed to overcome the difficulty encountered in solving the maximum weighted independent set~(MWIS) problem for wireless networks~\cite{DistributeSchedgnn}. The authors proposed a distributed MWIS solver based on spectral-based GCNs for link scheduling by combining the learning capabilities of GCNs and the efficiency of greedy MWIS solvers. The proposed solver achieves superior performance over greedy baselines with minimum increase in complexity, and generalizes well across different types of graphs and utility distributions.

Besides, the information of the dynamics, the topology structure and evolutionary patterns of dynamic networks can be fully exploited to improve the temporal link prediction performance. In dynamic network scenarios, K. Lei~\emph{et al.} introduced a novel non-linear GCN-GAN model by leveraging the benefits of spectral-based GCNs, LSTM as well as the GANs to tackle the challenging temporal link  prediction task~\cite{Gcn-gcnLP2019}. While in ultra-dense D2D mmWave communication networks, in order to effectively control the interference between communication pairs, spatial-based GCNs and primal-dual learning are utilized to solve the problem of joint beam selection and link activation across a set of communication pairs via inactivating part communication pairs~\cite{he2021gblinks}.

\subsubsection{Others} To improve the spectral efficiency in densely deployed WLANs, K. Nakashima~\emph{et al.} proposed a deep RL model using spectral-based GCNs for channel allocation~\cite{DeepRLCAgnn2020}. The idea behind their work is that the objective function is modeled as a parametric function of topologies, channels and communication quality. Z. He~\emph{et al.} studied the spectrum allocation via learning the low dimensional representations of a graph by modeling the V2X network as a graph, where each vehicle-to-vehicle link is a node in the graph~\cite{RAgnnVC2020}. According to the learned characteristics, multi-agent RL is used for spectrum allocation. DQN is used to learn to optimize the total capacity of the V2X network. In~\cite{ranasinghe2021graph}, GraphSAGE is employed to predict the potential links between APs for cell-free massive MIMO.

\subsubsection{Brief discussion} The application of GNNs in resource allocation mainly focuses on power allocation and link scheduling, while there is less work related to spectrum allocation and channel allocation. In addition, almost all works adopt GCNs and a few of them introduce RL in terms of the paradigms of GNNs. Although these works have achieved good results, there are also some issues needed to be considered. For one thing, resource allocation tasks in some wireless networks, such as intelligent factories and intelligent transportation, etc., may have many simple or complex constraints needed to be handled. Most of the existing works directly use projection strategy for simple constraints, while Lagrange dual learning framework is used for complex constraints. Although Lagrange dual learning framework can deal with complex constraints, it can not guarantee the complete satisfaction of constraints, and the training efficiency is not ideal. For another thing, in some wireless network scenarios with delay-sensitive traffics, the designed model should have low time complexity on the basis of achieving certain performance. A small amount of works has discussed the processing delay of the designed model, but there is still a certain distance from practical application. Therefore, the problem of model complexity is a direction worthy of exploration and research. On the other hand, most of the work is to solve the optimization problem of a single network layer. With the development of information and communication technologies, joint resource allocation at different network layers, i.e., cross-layer optimization, is a potential direction for the design of learning methods based on GNNs. The advantage of cross-layer optimization is that it can comprehensively utilize the information between different network layers, and then it may get performance improvement potentially.

\subsection{Emerging Fields}
GNNs are also applied in other wireless networks scenarios. Although these studies are in their infancy, the results of the existing works show that GNNs have a good application prospect in these directions.

\setcounter{subsubsection}{0}
\subsubsection{Channel estimation} Accurate CSI is important for guaranteeing the performance of massive MIMO high-dynamic networks. However, traditional solutions rely so much on hypothetical statistical models that they are hard to adapt the high-dynamic network environment. To overcome this defect, many researchers use DL to estimate channel of wireless communication system in recent years, but the ability of DL to extract spatial dependency is limited. However, GNNs have advantages in spatial dependency mining, so GNNs have been applied in this field in recent years.

The estimation of channels between the intelligent reflecting surface~(IRS), the base station~(BS), and the users is necessary for the optimal tuning of phase shifters at the IRS. J. Tang~\emph{et al.} proposed a DL model to configure the IRS and beamforming at the BS such that the system utility function is maximized directly based on the received pilots instead of the channel coefficients~\cite{BeaformIRS2021}. Specifically, spatial-based GCNs is utilized to directly map the received pilots to the beamformers at the BS and the reflective pattern at the IRS. While in~\cite{ChannelEstGAN}, GAT is employed to solve the channel estimation for the two-way backhaul link of high-altitude platform stations with reconfigurable intelligent surfaces. Numerical results show that for the full-duplex channel estimation, the performance of the GAT estimator is better than the least-squares. Moreover, numerical results also show that even if the training data does not include all changes, the GAT estimator is robust to hardware impairments and small-scale fading characteristics changes. As a further case of channel control, A. Scotti~\emph{et al.} considered the inference task of massive MIMO detection under time-varying channels and higher-order qadrature amplitude modulation and proposed a message-passing solution based on GNNs, i.e., MIMO-GNN~\cite{GnnMIMOdetect}.

Y. Yan~\emph{et al.} proposed a new channel tracking method based on GAE~\cite{GnnCTMIMO2020}. Specifically, the channel tracking framework is designed as
\begin{subequations}\label{Cachenable14}
\begin{align}
&\bar{e}_{i,j}^{'}(t) = \mathrm{MLP}_e^{coder}\left(\bar{e}_{i,j}(t), \bar{v}_i(t), \bar{v}_j(t)\right), \\
&\bar{v}_i^{'}(t) = \mathrm{MLP}_v^{coder}\left(\bar{v}_i(t), \sum\limits_{\bar{v}_j\in\mathcal{N}\left(\bar{v}_i(t)\right)}\bar{e}_{i,j}^{'}(t)\right), \\
&\hat{e}_{i,j}(t) = \mathrm{MLP}_e^{decoder}\left(\bar{e}_{i,j}^{'}(t)\right), \\
&\hat{v}_i(t)=\mathrm{MLP}_v^{decoder}\left(\bar{v}_i^{'}(t)\right),
\end{align}
\end{subequations}
where $\bar{v}_i(t)$ and $\bar{e}_{i,j}(n)$ are the resultant node feature of node $i$ and edge feature between nodes $i$ and $j$ at time $t$, respectively. $\mathrm{MLP}_v^{coder}$ and $\mathrm{MLP}_e^{coder}$ are coders for node and edge, respectively. In contrary, $\mathrm{MLP}_v^{decoder}$ and $\mathrm{MLP}_e^{decoder}$ are decoders. It's not hard to find that the designed framework combines MPNN and codec. Numerical results confirm that the GNN-based scheme outperforms the feed-forward neural network in terms of the MSE.

\subsubsection{Traffic prediction} Generally speaking, effective resource management can improve the utilization of network resources. In addition, if one can predict the required resource of future wireless traffic, resource management will become more flexible in wireless networks. However, the high spatial-temporal interdependencies make traffic prediction more challenging. There are fewer works on wireless network traffic prediction using GNNs, mainly including cellular network traffic prediction and satellite network traffic prediction.

\begin{figure}[t]
\renewcommand{\captionfont}{\footnotesize}
	\centering
\includegraphics[width=1.0\columnwidth,keepaspectratio]{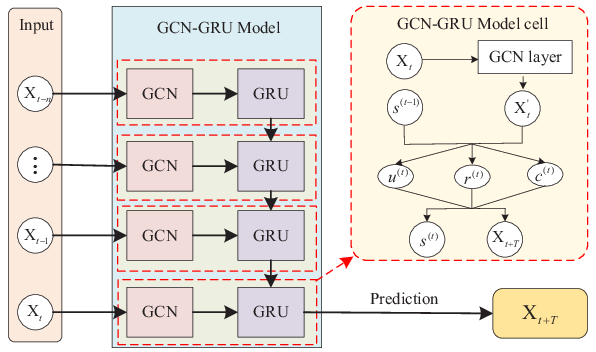}\\
	\caption{Illustration of GCN-GRU.}
	\label{GCN-GRU}
\end{figure}

To improve the accuracy of cellular traffic prediction, S. Zhao~\emph{et al.} proposed a new Spatio-Temporal GCNs incorporating Handover infOrmation~(STGCN-HO) prediction model using the transition probability matrix of the handover graph~\cite{CellNetTPgnn2020}. STGCN-HO builds a stacked residual neural network structure that combines spectral-based GCNs and CNN with gated linear units~\cite{GLU2017} to capture the spatial and temporal interdependencies of traffic. Compared with RNN, STGCN-HO has a faster training speed due to the use of CNN, and has the ability to train or predict cell or base stations with the information collected from the entire graph at the same time. In addition, compared with CNN grid, STGCN-HO can predict both base stations and the cells within the base stations. While in reference to the satellite network traffic prediction, L. Yang~\emph{et al.} pointed out that the traditional network traffic prediction model could not effectively extract the spatio-temporal characteristics of network traffic. Therefore, they proposed a network traffic prediction model GCN-GRU via combining the spectral-based GCNs with GRU~\cite{StlNetTPgcngru2020}. Specifically, GCN-GRU model utilizes the spectral-based GCN to extract the spatial characteristics of the satellite network traffic, and utilizes GRU model to extract the temporal characteristics of the satellite network traffic, and finally predict satellite network traffic through the fully connected layer. The illustration of GCN-GRU is shown in~Fig.~\ref{GCN-GRU}, where $r^{(t)}, u^{(t)}, c^{(t)}$ and $s^{(t)}$ correspond to \eqref{Cachenable8a}-\eqref{Cachenable8d} and $\mathrm{X}_t$ is the input feature at time $t$.

\subsubsection{Vehicle communication} 
\begin{figure}[t]
\renewcommand{\captionfont}{\footnotesize}
	\centering
\includegraphics[width=1.0\columnwidth,keepaspectratio]{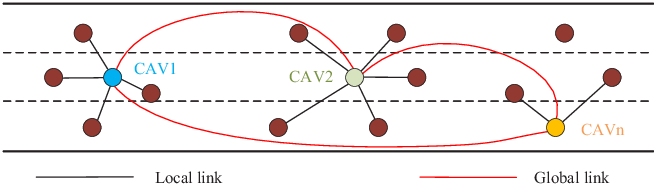}\\
	\caption{Illustration of graph construction for CAVs.}
	\label{CAV}
\end{figure}

Recently, GNNs have been applied to control the connected autonomous vehicles~(CAVs) lane changing decisions for a road segment, to mitigate the highway bottleneck congestion, and to allocate spectrum in V2X networks. J. Dong~\emph{et al.} proposed a DL model that combines spectral-based GCN and a deep Q network to control multiple CAVs to make cooperative lane change decisions~\cite{GCQNcavNet}. The graph construction methods of CAVs is shown in Fig.~\ref{CAV}. There is a state $S^{(t)}$ that is considered as a triplet at time $t$, i.e., $S^{(t)}=\{\mathbf{X}^{(t)}, \mathbf{A}^{(t)}, \mathbf{M}^{(t)}\}$, where $\mathbf{X}^{(t)}, \mathbf{A}^{(t)}$ and $\mathbf{M}^{(t)}$ denote the node feature matrix, adjacent matrix and a mask matrix that document the index of autonomous vehicles at time $t$, respectively. From the perspective of CAV operations, the proposed model not only enables CAV to successfully carry out lane changes to meet its personal intention of merging from the prescribed ramp, but also guarantees safety and efficiency. Similarly, RL algorithms are employed to train CAV driving behaviors, which can be used to relieve highway bottleneck congestion~\cite{cavMARL}.

\subsubsection{Others} In addition to the aforementioned related works, GNNs are also used to solve other problems in wireless networks. J. Shao~\emph{et al.} proposed Branchy-GNN using the branch network and source-channel coding to reduce the computational cost and intermediate feature transmission overhead for efficient point cloud processing~\cite{Branchy-GNN}. T. Fujihashi~\emph{et al.} proposed a novel soft point cloud transmission scheme that combines GNN-based point cloud coding and near-analog modulation for future wireless streaming of holographic and three-dimensional data~\cite{Wireless3Dpcgnn}. C. S. Jeong \emph{et al.} provided AR services via three-dimensional GNN using cameras and sensors on mobile devices~\cite{ARanchor2021}. S. Lohani~\emph{et al.} designed a model based on spectral-based GCNs to maximize the throughput of unmanned aerial vehicle~(UAV)-assisted ground networks~\cite{FSOAMC2019}. Spectral-based GCNs were utilized to solve the network localization problem of a wireless network in two-dimensional space~\cite{NetworkLoc2021}. In~\cite{soto2021atari}, spatial-based GCNs was first used to predict the achieved throughput in highly dense WLANs using channel bonding. Using the topology extraction ability of GNN, spectral-based GCNs based learning routing scheme was proposed to implement onboard routing in small satellite networks~\cite{liu2021routing}. M. Lee \emph{et al.} analyzed and enhanced the robustness of the decentralized GNN in different wireless communication systems, making the prediction results not only accurate but also robust to transmission errors~\cite{DecentralizedInfer2021}.

\subsubsection{Brief discussion} Emerging fields where GNNs are applied mainly include channel estimation, channel tracking, MIMO detection, traffic prediction, vehicle communication, point clouds, and so on. There is not much work in these fields, but some fields are worthy of further exploration. For instance, the traffic prediction task not only plays a pivotal role in the management of wireless network resources but also has higher requirements for data collection and acquisition. On the one hand, traffic prediction generally requires long-term historical data. It also involves issues such as the granularity of data collection, the level of data collection, and data privacy, which in turn brings varying degrees of difficulties. On the other hand, for scenarios with high requirements for traffic prediction service delay, the model designed is required to be as low as possible in complexity, so as to achieve real-time prediction capabilities. In addition, in terms of the application of the paradigms of GNNs in emerging fields, GCNs are frequently utilized. While there are also a small number of new paradigms of GNNs applied, such as GAT, GAE, and several generalized GCNs. According to the characteristics of different business needs, we can measure and compare different GNNs paradigms, and then adopt the best GNNs paradigm to better solve the problems faced.

\section{Key Issues and Future Development}
Although GNNs have made some progress in the application of wireless networks, some key issues need to be further studied in-depth. Accordingly, some ongoing or future research directions that are worth exploring are summarized as follows:

\setcounter{subsubsection}{0}
\subsubsection{Acquisition of high-quality data}One of the fundamental elements of the data-driven DL method is to obtain a lot of training and testing data of wireless networks. The higher the quality of data, the better the training of the model. Though many researchers have studied the application of DL in wireless networks, most of the datasets used in the existing work are generated by numerical simulation, which is somewhat different from the real data of wireless networks. Furthermore, unlike the successful application of DL in image processing and social networks, etc, there is a lack of publicly recognized data set for the physical layer and media access layer of wireless networks. Therefore, it is urgent to collect and construct opening wireless network datasets for method comparison and performance verification.

Wireless networks have some unique characteristics, such as high dynamic, heterogeneous terminals, and non-uniform, resulting in many difficulties in obtaining the data of wireless networks, especially the real-time communication data of the physical layer and media access control layer. The data of wireless networks can generally be obtained from spectrum measurement instruments, base station, core network equipment, user terminals, and so on. Different data acquisition devices may be provided by different manufacturers, who may define different data extraction formats, feature names, and data calculation methods, and may also be different in the time granularity of extraction. This makes it very challenging to collect massive data in the real wireless network, especially at the physical layer and media access control layer. Meanwhile, the construction of opening testing and training datasets is still a very urgent and challenging task for the successful application of DL and GNN, etc, in wireless networks.

\subsubsection{Distributed GNN learning model}As we all know, the goal of whether the data-driven DL methods or the data- and model-driven methods is to learn the super parameters of NNs that depend on the supercomputing power. Furthermore, the learning abilities of the DL methods are proportional to the network complexity. However, in wireless networks, the computing capacity of communication nodes is very limited, especially for the battery-powered lightweight devices. How to design a proper DL model is a challenging and opening problem for the battery-powered lightweight devices. Fortunately, distributed ML is regarded as an effective and efficient technology to balance the performance and the computational resource and to reduce the required amount of training and testing data. However, several issues are needed to be considered when designing a GNN-based distributed learning model, such as the split of the GNN-based learning model, the parameter updating strategy, the integration of the results of each distributed running node, etc. In addition, the convergence of the whole distributed learning model is also should be guaranteed.

\subsubsection{Data privacy issues}Generally speaking, the acquisition of the data of wireless networks inevitably touches the user's privacy. However, privacy protection is one of the core issues in the field of information and communications technology in the future, and the process of data acquisition is cumbersome. Consequently, the acquisition of real network data will face many difficulties. On the other hand, distributed learning model needs data interaction, which also involves data privacy. In some scenarios with high requirements for data privacy protection, data interaction security needs to be ensured. Some methods can be used to solve data privacy issues, such as secure multi-party computing and trusted execution environments, etc. Another popular solution is to use the federated learning, which has the outstanding advantage of privacy protection.

\subsubsection{Robustness of the learning model}Although DL methods have been successfully applied in many fields, their effectiveness depends on high-quality training datasets. When the training dataset contains significant complex noise, anomaly intrusion, category imbalance, etc., its effectiveness cannot be guaranteed. This implies that the robustness of the learning model should be considered when designing the DL model for wireless networks. Generally, the robustness of the model can be discussed from two aspects, i.e., malicious or abnormal data and the defense ability of the model. On the one hand, when a malicious attacker attacks the model with input data, it can generally be defended by adversarial training, input transformation, gradient shielding, and detection and rejection. On the other hand, the model's ability to resist attacks can be improved by improving the model itself and combining it with other security technologies.

\subsubsection{Scalability of the learning model}In wireless networks, network scale, network state, and network service data volume may be highly dynamic, which requires stronger scalability of the designed model. For example, the changes in network scale (such as changes in the number of communication devices) and network state (such as changes in the location and attributes of network communication nodes) are enough to bring great challenges to the design of the learning model. Fortunately, GNNs have the ability to address these problems based on the existing research work, but the current work is still in its infancy and needs further research. On the other hand, with the high dynamic change of network business data volume, such as the sharp increase of business data volume, the real-time processing capability of the model should be enhanced accordingly.

\subsubsection{Cross-layer optimization} Most of the existing works have covered different individual network layers in terms of GNNs applications. These algorithms generally cannot obtain the global optimal solution from the perspective of whole communication networks. More recently, with the development of information and communication technologies, joint resource allocation at different network layers, i.e., cross-layer optimization, is regarded as a potential direction for further improving the performance of learning methods. The advantage of cross-layer optimization is that the information at multiple network layers can be fully explored and exploited to design learning methods. Although cross-layer optimization has the potential to improve performance, it introduces more optimization parameters. Meanwhile, the objective function will be more complicated, which in turn requires higher computing power and may introduce relatively large processing delays. Parameter abstraction can reduce the complexity to a certain extent, but it may reduce the optimality of the generated configuration.

\subsubsection{Research of GNN-based deep unfolding method}One shortcoming of the data-driven learning method is the lack of interpretability due to the NN-based methods treat the learning of mapping between the input and output as a black box, which weakens the domain knowledge. To tackle the shortcoming, an emerging direction is the algorithm unfolding (unrolling) aiming at combining the knowledge of data and domain fields. Motivated by the successful application of algorithm unfolding solving some classical problems like UWMMSE, recently, deep unfolding is regarded as an effective combination to not only effectively utilize the interpretability and scalability of model-driven algorithms, but also use the expressive power of data-driven methods. Although deep unfolding has been paid a lot of attention from both the industry and academia, the problem to be solved has only a few simple constraints. Meanwhile, there is a little work to deeply study the GNN-based unfolding models to solve the optimization problem in the wireless network under practical constraints. This implies that there are still a lot of works to be explored for wireless networks.

\subsubsection{Model design with complex constraints}So far, a large amount of work focuses on designing data-driven or data- and model-driven DL models aiming to solve the optimization problem of wireless network, however, these problem investigated is generally formulated subjecting to simpler constraints. In practical wireless communication, the transmission schemes may be investigated under some complex constraints, such as the per-antenna or per-base station power constraint, the quality of service constraint, delivery latency, etc. How to design an effective and efficient DL model under complex constraints is still an opening and challenging issue. Although Lagrange dual learning framework has been used to deal with such problems, there are still some problems with this method. On the one hand, the convergence speed and the effectiveness of the Lagrange dual learning framework are closely related to the updating step-sizes of the Lagrange multipliers and the model parameters, while the update step-sizes of the Lagrange multipliers and the model parameters are coupled with each other, which makes it difficult to determine the approximate update step-sizes to maximize the performance of the DL model. On the other hand, the DL models trained based on Lagrange dual learning framework are generally difficult to fully satisfy constraints (especially for the problems with large-scale complex constraints), and the generalization performance of models is also questionable.

\section{Conclusions}
In this paper, we first illustrated the construction method of WCG for various wireless networks. Then, we simply introduced several classical paradigms of GNNs that have been applied in wireless networks, and made a classified introduction for the application of GNNs in wireless networks, mainly including resource allocation and several emerging fields. From the overview results, the application of GNNs in wireless networks is still in its infancy. Many challenging problems are needed to be further solved and improved. Finally, based on the existing results, several key issues and research directions are summarized for participators interested in this domain.

\bibliographystyle{IEEEtran}
\bibliography{reference}

\begin{IEEEbiography}[{\includegraphics[width=1in,height=1.25in,clip,keepaspectratio]{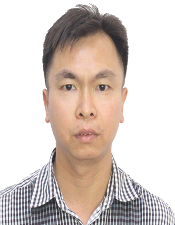}}]{Shiwen He} (M'14) received the M.S. degree in the computational mathematics from Chengdu University of Technology, Chengdu, China, and the Ph.D. degree in the information and communication engineering from Southeast University, Nanjing, China, in 2009 and 2013, respectively.

Since April 2018, he has been a Professor with the School of Computer Science and Engineering, Central South University, Changsha, China. From November 2015 to March 2018, he has been an Associate Research Fellow with the School of Information Science and Engineering, Southeast University. From September 2013 to October 2015, he was a Postdoctoral Researcher with the State Key Laboratory of Millimeter Waves, Department of Radio Engineering, Southeast University. He has authored or coauthored more than 125 technical publications and held 105 invention patents granted. His research interests include wireless communications and networking, distributed learning and optimization computation theory, and big data analytics.

Dr. He has been a technical Associate Editor for the IEEE 802.11aj from December 2015 to July 2018 and submitted more than 20 technical proposals to IEEE standards. He was the recipient of the Contribution Award for the development of the IEEE 802.11aj Standard published on April 18, 2018. He has served as an associate editor for IEEE Signal Processing Letters and EURASIP Journal on Wireless Communications and has been a TPC member of various conferences, including Globecom, ICC, ICCC, and WCSP, etc.
\end{IEEEbiography}
\begin{IEEEbiography}[{\includegraphics[width=1in,height=1.25in,clip,keepaspectratio]{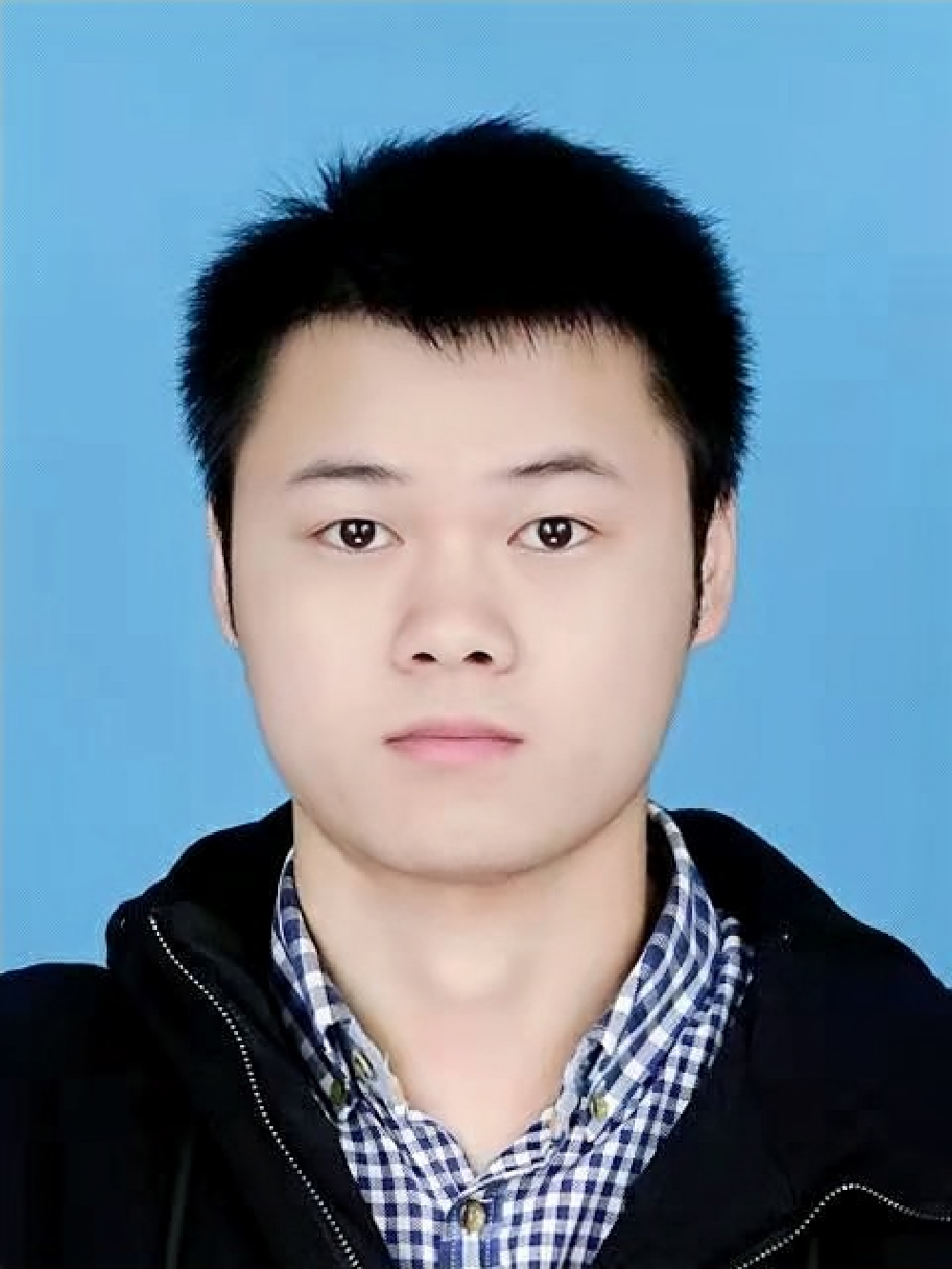}}]{Shaowen Xiong} received the B.S. degree in information and computing sciences from the Hunan University of Technology, Zhuzhou, China, in 2019. He is currently pursuing the M.S. degree in computer science and technology from the Central South University, Changsha, China. His current research interests include graph neural networks and its application on wireless communication networks.
\end{IEEEbiography}
\begin{IEEEbiography}[{\includegraphics[width=1in,height=1.25in,clip,keepaspectratio]{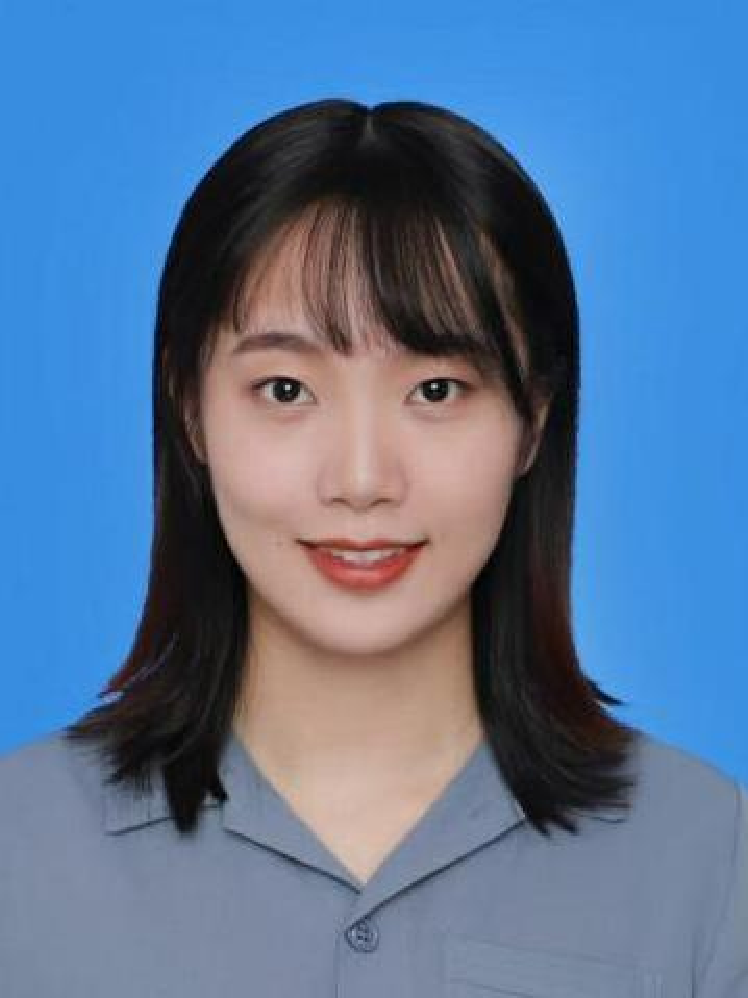}}]{Yeyu Ou} received the B.S. degree in engineering mechanics from the Hehai university, Nanjing, China, in 2020. She is currently pursuing the M.S. degree in computer technology from the Central South University, Changsha, China. Her current research interests include graph neural networks and big data analysis.
\end{IEEEbiography}

\begin{IEEEbiography}[{\includegraphics[width=1in,height=1.25in,clip,keepaspectratio]{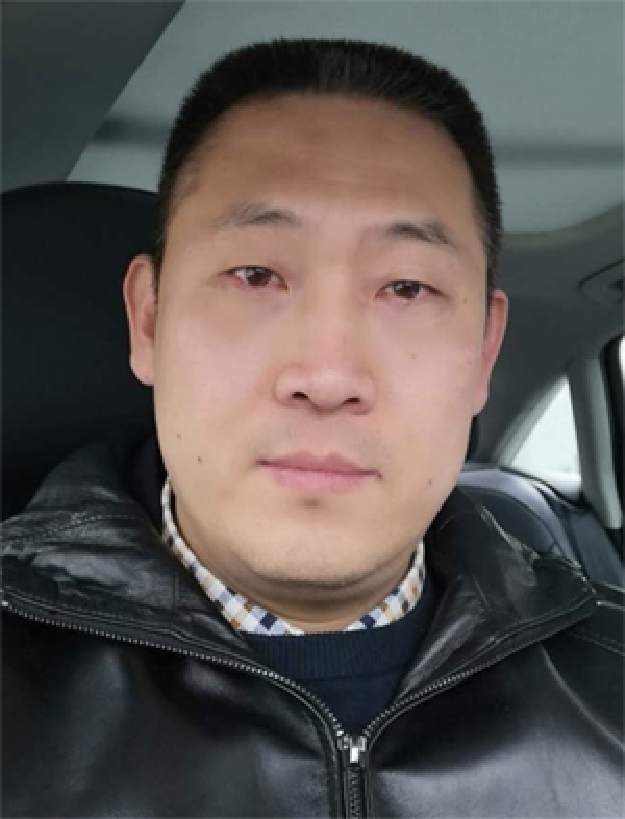}}]{Jian Zhang}(M'19) received the B.Eng. degree in computer science from the National University of Defense Technology, in 1998, and the M.Eng. and Ph.D. degree in computer science from Central South University, in 2002 and 2007, respectively, where he is currently an Associate Professor with the School of Computer Science and Engineering.

His research interests include optimization theory, cyberspace security, cloud computing, and cognitive radio technology. He has published over 30 peer-viewed journal papers, and is holding 12 granted and filed patents.
\end{IEEEbiography}

\begin{IEEEbiography}[{\includegraphics[width=1in,height=1.25in,clip,keepaspectratio]{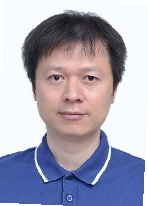}}]{Jiaheng Wang} (M'10-SM'14) received the Ph.D. degree in electronic and computer engineering from the Hong Kong University of Science and Technology, Kowloon, Hong Kong, in 2010, and the B.E. and M.S. degrees from the Southeast University, Nanjing, China, in 2001 and 2006, respectively.

He is currently a Full Professor at the National Mobile Communications Research Laboratory (NCRL), Southeast University, Nanjing, China. From 2010 to 2011, he was with the Signal Processing Laboratory, KTH Royal Institute of Technology, Stockholm, Sweden. He also held visiting positions at The Friedrich-alexander University of Erlangen-nurnberg, Germany, and the University of Macau, Macau.

His research interests are mainly on wireless communications and networks. Dr. Wang has published more than 100 articles on international journals and conferences. From 2014 to 2018, he served as an Associate Editor for the IEEE Signal Processing Letters. From 2018, he serves as a Senior Area Editor for the IEEE Signal Processing Letters. He is a recipient of the Humboldt Fellowship for Experienced Researchers.
\end{IEEEbiography}

\begin{IEEEbiography}[{\includegraphics[width=1in,height=1.25in,clip,keepaspectratio]{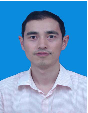}}]{Yongming Huang} (M'10-SM'17) received the B.S. and M.S. degrees from Nanjing University, China, in 2000 and 2003, respectively. In 2007 he received the Ph.D. degree in electrical engineering from Southeast University, China.

Since March 2007 he has been a faculty in the School of Information Science and Engineering, Southeast University, China, where he is currently a full professor. During 2008-2009, Dr. Huang was visiting the Signal Processing Lab, Electrical Engineering, Royal Institute of Technology (KTH), Stockholm, Sweden. His current research interests include MIMO wireless communications, cooperative wireless communications and millimeter wave wireless communications. He has published over 200 peer-reviewed papers, hold over 60 invention patents. He submitted around 20 technical contributions to IEEE standards, and was awarded a certificate of appreciation for outstanding contribution to the development of IEEE standard 802.11aj. He has served as an Associate Editor for the IEEE Transactions on Signal Processing, and a Guest Editor for the IEEE Journal Selected Areas in Communications, is serving as an Editor-at-Large for the IEEE Open Journal of the Communications Society, and an Associate Editor for the IEEE Wireless Communications Letters.
\end{IEEEbiography}

\begin{IEEEbiography}[{\includegraphics[width=1in,height=1.25in,clip,keepaspectratio]{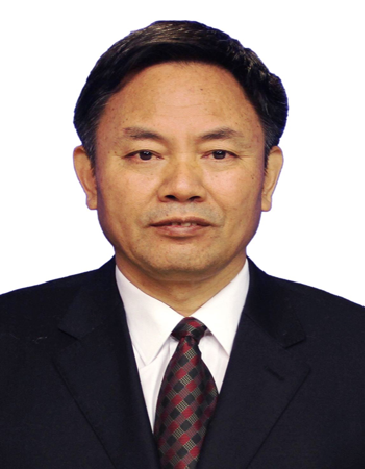}}]{Yaoxue Zhang} [M'17-SM'18] received his B.Sc. degree from Northwest Institute of Telecommunication Engineering, China, in 1982, and his Ph.D. degree in computer networking from Tohoku University, Japan, in 1989. Currently, he is a professor with the School of Computer Science and Engineering, Central South University, China, and also a professor with the Department of Computer Science and Technology, Tsinghua University, China. His research interests include computer networking, operating systems, ubiquitous/pervasive computing, transparent computing, and big data.

He has published over 200 technical papers in international journals and conferences, as well as 9 monographs and text- books. Currently, he is serving as the Editor-in-Chief of Chinese Journal of Electronics. He is a fellow of the Chinese Academy of Engineering.
\end{IEEEbiography}

\end{document}